\documentclass{elsart}

\usepackage{graphics}

\usepackage{amsmath}
\usepackage{amsfonts}
\usepackage{amssymb}
\usepackage{natbib}

\journal{Bulletin of Mathematical Biology}

\renewcommand{\exp}[1]{\e^{#1}}
\renewcommand{\vec}[1]{\mathbf{#1}}
\renewcommand{\d}{\mathrm{d}}
\newcommand{\order}{\mathcal{O}}

\begin{document}

\begin{frontmatter}

\title{A nonlocal continuum model \\ for biological aggregation}

\author[UCLA]{Chad M. Topaz\corauthref{cor}},
\corauth[cor]{Corresponding author.}
\ead{topaz@ucla.edu}
\author[UCLA]{Andrea L. Bertozzi},
\author[UAB]{Mark A. Lewis}
\address[UCLA]{Department of Mathematics \\ UCLA, Los Angeles, CA, 90095, USA}
\address[UAB]{Department of Mathematical and Statistical Sciences \\ Department of Biological Sciences \\  University of Alberta, Edmonton, AB, T6G 2G1, Canada}

\begin{abstract}
We construct a continuum model for biological aggregations in which individuals experience long-range social attraction and short range dispersal. For the case of one spatial dimension, we study the steady states analytically and numerically. There exist strongly nonlinear states with compact support and steep edges that correspond to localized biological aggregations, or clumps. These steady state clumps are approached through a dynamic coarsening process. In the limit of large population size, the clumps approach a constant density swarm with abrupt edges. We use energy arguments to understand the nonlinear selection of clump solutions, and to predict the internal density in the large population limit. The energy result holds in higher dimensions, as well, and is demonstrated via numerical simulations in two dimensions.
\end{abstract}

\end{frontmatter}


\section{Introduction}

Biological aggregations such as insect swarms, ungulate herds, fish schools, and bacterial colonies are widespread examples of self-organization in nature \citep{ph1997,bch2000,ol2001,cdfstb2001}. These groups often arise as social phenomena, without direction from a leader or influence of external stimuli such as food and light sources. Social forces among organisms include attraction, for group cohesion, and repulsion, for collision avoidance \citep{b1954,mk1999}. The resulting aggregations can confer benefits such as protection and mate choice to their members \citep{pk1999}.

Mathematical models of social aggregations can be classified into Lagrangian and Eulerian types. The Lagrangian approach treats each organism as a particle obeying a nonlinear difference or differential equation \citep{s1973b,ss1973,osik1977,vcbcs1995,lrc2001,set2001,ckjrf2002,eea2002,pvg2003,ah2003,ee2003,mkbs2003}. Alternatively, the Eulerian approach describes the local flux of individuals with an advection-diffusion equation for a continuum population density field \citep{k1978,o1980,my1982,dd1984,i1984,a1985,i1985,sm1985,in1987,hm1989,go1994,kwg1998,tt1998,fglo1999,mk1999,tb2004}. A variety of methods can be used to connect the two formulations.  The usual method involves a Fokker-Planck approximation which relates the distribution of jump distances made by individuals to terms in the advection-diffusion equation \citep{ol2001}. Since the social communications between organisms often take place at large distances via sight, sound, or smell, models may be nonlocal in space \citep{k1978,a1985,i1985,sm1985,in1987,hm1989,go1994,fglo1999,mk1999,ogk2001,tb2004}.

Biological aggregations can form distinct groups with sharp edges as they move \citep{ph1997,pk1999}.
We refer to this phenomenon as clumping. While clumping has been observed in two-dimensional numerical numerical \citep{lrc2001} and analytical \citep{tb2004} models, recent mathematical analyses of swarming behaviors \citep{mk1999} were unable to find the same kind of clumping from biologically reasonable one-dimensional models.  As we will describe below, earlier one-dimensional models which support clumping behavior include assumptions about biological interactions which are unlikely to be met in nature.

Our goal in this paper is to investigate how clumping can arise in a simple, realistic nonlocal model for biological aggregation, both in one spatial dimension and in higher dimensions. Our Eulerian model is 
an integrodifferential conservation law with two movement terms. One describes nonlinear degenerate diffusion arising from anti-crowding behavior, and the other describes attractive nonlocal social interactions. In contrast to previous studies \citep{my1982,nm1983,kwg1998,mk1999}, we do not seek traveling solutions, which would correspond to cohesive group movement. Rather, we are concerned with stationary solutions. Our formulation relates to several earlier models that have appeared in the mathematical literature. We discuss these now, and then state our main results and outline the remainder of this paper.

Many nonlocal continuum models for biological aggregation may be cast in the form
\begin{equation}
\frac{\partial\rho}{\partial t}=\frac{\partial}{\partial x}
\left(D\frac{\partial \rho}{\partial x}-v_a\rho\right).
\label{eq:early}
\end{equation}
Here $x$ is the one-dimensional space coordinate, $t$ is time, $\rho(x,t)$ is the local population density, $D$ measures diffusion, and $v_a$ is the nonlocal, density-dependent speed.

The earliest models take $D$ constant, and $v_a=K*u$. The asterisk denotes spatial convolution and $K$ is an odd spatial weighting function which drops off with distance and has finite mass \citep{k1978,go1994}. $K$ models attractive social forces, which can give rise to spatial instabilities leading to a unique steady state with a non-uniform spatial distribution of the population~\citep{go1994}.  However, these patterns do not include clumps (aggregations with compact support).

Extensions by \citet{mk1999} consider group drift by including a local, density-dependent velocity as well as by including an even component in $K$, giving rise to traveling swarms. These swarms are not stable over long periods of time, having a tendency to break down by losing individuals at the rear of the swarm \citep{mk1999}. However, the authors note (via analysis and numerical experiments) that nonlinear diffusion has a tendency to stabilize the swarms.

On the other hand, earlier extensions of the work of \citet{k1978} by \citet{my1982}, \citet{nm1983}, \citet{a1985}, \citet{i1985} and \citet{in1987} include density dependent (rather than constant) diffusion, still coupled to long-range advective attraction. The model of \citet{a1985} is inspired by chemotactic locomotion, and includes nonlinear diffusion which is degenerate for finite $\rho$. In contrast, in the work of \citet{my1982}, \citet{nm1983}, \citet{i1985} and \citet{in1987}, the diffusion in (\ref{eq:early}) takes the form $D = p\rho^{p-1}$. The parameter $p$ determines the degree of nonlinearity of the diffusion, and may be freely varied. In all of these works, the aggregative weighting function $K$ does not meet our previous assumptions of having a finite mass and decaying with distance. Further, it has support $\ell$ which is either a tunable parameter \cite{i1985,in1987} or is infinite \citep{my1982,nm1983}. The model of \citet{hm1989} is similar, but adds reaction-type terms modeling logistic growth and predation. Depending on $p$ and $\ell$, the nonlinear diffusion models have stationary \citep{i1985,in1987,hm1989} and traveling \citep{my1982,nm1983} clump solutions.  Although this class of models exhibits the desirable behavior of clump formation, its limitation arises from the biologically unrealistic assumption of strong attractive interactions between individuals over arbitrarily large distances.

In this paper we modify the classic model of \citet{k1978} to include density-dependent diffusion.  We show that this modification is sufficient to give rise to clumped solutions with very sharp edges. These solutions can be understood using classical applied mathematical methods of weakly nonlinear analysis, phase plane analysis, and energy methods. Furthermore, using asymptotic and scaling arguments, we show how, in the large population size limit, the clumps have constant internal population density. The preferred density is predicted by analyzing the energy. The constant internal density property is typical of biological groups (see \citeauthor{pk1999}, \citeyear{pk1999} and the extensive discussion in \citeauthor{mkbs2003}, \citeyear{mkbs2003}) but has not been found in other realistic continuum models. Our model has this property not only in one spatial dimension, but in higher dimensions as well.

The remainder of this paper is organized as follows. We formulate our model in Section \ref{sec:model}. In Section \ref{sec:characteristics}, we outline its basic features, including its conservation properties, linear stability, and energy. In Section \ref{sec:1dexample}, we pose the model on a periodic one-dimensional domain and perform an in-depth study for a particular choice of the aggregative weighting function $K$. We analyze instabilities of trivial population profiles using weakly nonlinear methods. We then use phase plane methods to deduce the qualitative structure of swarm patterns. We extend the analysis of the qualitative structure through the numerical calculation of nonlinear steady-state solutions, and the construction of numerical bifurcation diagrams. Interestingly, for fixed population size, the allowed clumps solutions are not unique. We use energy methods discover which stationary solutions are preferred, allowing us to deduce the pattern selection mechanisms that drive the dynamical system.  We also investigate the effects of domain size and population size, and perform numerical simulations which display the model's coarsening behavior. Section \ref{sec:2d} contains a brief extension in which we perform numerical simulations in two spatial dimensions to show clump formation. Finally, we summarize and conclude in Section \ref{sec:conclusion}.

\section{Mathematical model}
\label{sec:model}

Consider a population density $\rho(\vec{x},t) \geq 0$ which moves with velocity $\vec{v}(\vec{x},t)$, $\vec{x},\vec{v} \in \mathbb{R}^n$, $t \geq 0$. We assume that birth, death, immigration, and emigration of organisms are negligible on the time scale of interest. Then $\rho$ satisfies the standard conservation equation
\begin{equation}
\label{eq:cd}
\rho_t + \nabla \cdot (\vec{v} \rho) = 0.
\end{equation}
The members of the population move towards and away from each other following basic biological principles of aggregation and dispersal, \emph{i.e.}
\begin{equation}
\label{eq:v}
\vec{v} = \vec{v}_a + \vec{v}_d.
\end{equation}
We now develop a kinematic model  \citep{kwg1998,mk1999,tb2004} which assumes that the attractive and dispersive velocities $\vec{v}_a$ and $\vec{v}_d$ depend only on properties of $\rho$ at the current time.

Though the particulars of aggregation vary from species to species, the sensing mechanism responsible  (\emph{e.g.} sight or smell) typically has a characteristic range, which we call $\ell$, and degrades over distance. For simplicity, we assume that the sensing is spatially isotropic, and that organisms sense an averaged nearby population. We then associate with an individual in the population at position $\vec{x}$ the sensing function
\begin{equation}
\label{eq:sensing}
s(\vec{x}) = \int_{\mathbb{R}^n} K(\vec{x} - \vec{y}) \rho(\vec{y})\ d\vec{y}  \equiv K * \rho.
\end{equation}
The  $*$ operator denotes convolution. The kernel $K$ incorporates the sensing range and degradation for the particular species under consideration. We assume, without loss of generality, that $K$ is normalized such that $\int_{\mathbb{R}^n} K(\vec{x})\ d\vec{x} = \ell^n$. Individuals aggregate by climbing gradients of the sensing function $s(\vec{x})$. We denote the species-specific characteristic attractive movement speed by $V$. The attractive velocity, then, is
\begin{equation}
\label{eq:va}
\vec{v}_a = \frac{V \ell}{\alpha} \nabla(K * \rho).
\end{equation}
Here, $\alpha$ is a characteristic density. Dimensional arguments dictate that the factor of $\ell/\alpha$ must appear, so that $\vec{v}_a$ has the correct units of velocity. Since density has units of number of organisms per unit space, we form the characteristic density
\begin{equation}
\alpha = \frac{1}{\ell^n}. \label{eq:alpha}
\end{equation}
That is to say, we take the characteristic density to be the one for which the spacing between organisms is the characteristic sensing range.

Dispersal is assumed to arise as an anti-crowding mechanism, and operates over a much shorter length scale \citep{b1954,mk1999}. Correspondingly, we take it to be spatially local and in the opposite direction of population gradients.  Moreover, anti-crowding motion is assumed to decrease as the population thins. The simplest model describing these effects is linear in both $\rho$ and $\nabla \rho$, \emph{i.e.}
\begin{equation}
\label{eq:vd}
\vec{v}_d = - \frac{V r \ell}{\alpha^2} \rho \nabla \rho
\end{equation}
Here, $r$ quantifies the ratio of typical aggregative to repulsive velocities. The factor of $\ell/\alpha^2$ arises from dimensional arguments, similar to those above for the aggregative velocity.

Combining (\ref{eq:cd}), (\ref{eq:v}), (\ref{eq:va}), (\ref{eq:alpha}) and (\ref{eq:vd}), we obtain an integrodifferential equation for $\rho$, namely
\begin{equation}
\label{eq:gedimensional}
\rho_t + \nabla \cdot \left(V \ell^{n+1} \rho K * \nabla \rho - V r \ell^{2n+1} \rho^2 \nabla \rho \right) = 0
\end{equation}
where we have used the commutativity of convolution and differentiation. This model is related to the one-dimensional models proposed by \citet{my1982}, \citet{nm1983}, \citet{i1985} and \citet{in1987}  (those models may be put in the same form as ours by using commutativity of  differentiation and convolution). In those models, the dispersal rate is a (tunable) power of the density, and the sensing kernels $K$ \emph{grow} linearly in space with a possible cutoff at a finite range \citep{i1985,in1987}. Our model is also similar to that in \citet{mk1999}. However, there, the aggregation and repulsion effects depend directly on the same sensing function. Our goal in this paper is to present a model that is valid in higher dimensions and incorporates biologically reasonable assumptions about aggregation and dispersal, in particular that the former occurs on a longer scale than the latter.

We rescale the variables as $\widetilde{x} = (1/\ell)x$, $\widetilde{t} = (V/\ell)t$, $\widetilde{\rho} = (1/\alpha) \rho $ and let $\widetilde{K}(\widetilde{x}) = K(x)$. Substituting into (\ref{eq:gedimensional}) and dropping the tildes for convenience, we arrive at the dimensionless governing equation
\begin{equation}
\label{eq:ge}
\rho_t + \nabla \cdot (\rho K * \nabla \rho - r \rho^2 \nabla \rho ) = 0.
\end{equation}
Due to our rescaling, the characteristic distance of attraction in $K$ is now one, and $\int_{\mathbb{R}^n} K(\vec{x})\ d\vec{x} = 1$.

As previously discussed, $K$ should reflect that biological sensory mechanisms have limited spatial extent. Common choices \citep{mk1999,ogk2001} include the decaying exponential and the characteristic function:
\begin{subequations}
\begin{eqnarray}
\label{eq:k}
K^e_n(\vec{x}) & \equiv & \alpha_n \e^{-|\vec{x}|}, \quad  \quad \alpha_n = \begin{cases} 1/2, &\text{$n=1$}\\ 1/(2\pi), &\text{$n=2$}\\ 1/(4\pi^2), &\text{$n=3$} \end{cases} \label{eq:kexp} \\
K^c_n(\vec{\vec{x}}) & \equiv & (1/2^n)\chi_{S^n}, \quad  S^n = \begin{cases} [-1,1], &\text{$n=1$}\\ [-1,1] \times [-1,1],&\text{$n=2$}\\ [-1,1] \times [-1,1] \times [-1,1],&\text{$n=3$} \end{cases}.
\end{eqnarray}
\end{subequations}
The exponential function $K^e_n$ arises from assuming a constant rate of transmission failure of sensory data per unit distance (\emph{i.e.} the constant hazard function assumption). More complex behavioral models for $K^e_n$ also include differential weighting given to stimuli received from different distances. The characteristic function $K^c_n$, on the other hand, arises under the assumption of an identical response to all individuals within a fixed distance. 

In order to fully define the model, we must specify a domain $D$ along with suitable boundary conditions for (\ref{eq:ge}). We consider three choices. The first is no flux boundary conditions, which implies that the members of population cannot enter or leave the domain. The second choice is periodic boundary conditions, which are mathematically convenient. For this case, the kernel $K$ must be modified to be periodic in order for the model to make sense. The final choice is to solve the problem in free space, \emph{i.e.} $D=\mathbb{R}^n$, which corresponds to a biological population with no nearby physical barriers. In Section~\ref{sec:1dexample} we present a case study in one spatial dimension with periodic boundary conditions. In that section, we discuss how this choice also includes the case of no flux boundary conditions, and how it may be used to approximate the free space problem.

\section{Basic model characteristics}
\label{sec:characteristics}

The $n$-dimensional model (\ref{eq:ge}) has a number of useful general properties which we describe in this section. In particular, we discuss conservation properties of the model and its linear stability. We also show the existence of a Lyapunov functional (or energy) which is dissipated under the dynamics of the governing equation (\ref{eq:ge}). This is, to our knowledge, the first time such an energy has been introduced. We use the energy later to understand the selection of nonlinear swarm states, and to obtain a quantitative prediction of the equilibrium population density for large swarms.

\textbf{Conservation of moments.} We first discuss the model's conservation properties (temporarily restricting attention to the free space problem, \emph{i.e.} $D=\mathbb{R}^n$). By construction (see Section \ref{sec:model}) (\ref{eq:ge}) conserves the zeroth moment of $\rho$, namely the total population size or \emph{mass} $M = \int_{D}\rho\ d\vec{x}$. Equation (\ref{eq:ge}) also conserves the first moment, or scaled center of mass, $M^{(1)} = \int_D\vec{x} \rho\ d\vec{x}$. For convenience, we define the inner product $(\vec{a},\vec{b}) =\int_D \vec{a} \cdot \vec{b}\ d\vec{x}$, where the dot product must be interpreted appropriately in the case that $\vec{a}$ and/or $\vec{b}$ are scalars. To show that $M^{(1)}$ is conserved, we consider
\begin{subequations}
\begin{eqnarray}
M^{(1)}_t & = & (\rho_t,\vec{x}) \\
& = & (-\nabla \cdot [\rho K*\nabla \rho - r \rho^2 \nabla \rho],\vec{x}) \label{eq:momentone1} \\
& = & (\rho K*\nabla \rho - r \rho^2 \nabla \rho,1) \label{eq:momentone2} \\
& = & (\rho K*\nabla \rho,1) - (r \rho^2 \nabla \rho,1) \label{eq:momentone3} \\
& = & (\rho K*\nabla \rho,1) - (r \nabla \rho^3/3,1) \label{eq:momentone4} \\
& = & (K*\nabla \rho,\rho) - 0 \label{eq:momentone5}
\end{eqnarray}
\end{subequations}
where we have assumed that there is no population contained at infinity. Equation (\ref{eq:momentone1}) comes from substituting (\ref{eq:ge}), and (\ref{eq:momentone2}) and (\ref{eq:momentone5}) come from integration by parts. We then note that $(K*\nabla \rho,\rho) = (\nabla[K*\rho],\rho) = -(K*\rho,\nabla \rho)$ by commutativity of differentiation and convolution, and integration by parts. Also, we have that $(K*\nabla \rho,\rho) = (\nabla \rho,K*\rho) = (K*\rho,\nabla \rho)$ since the convolution may be moved across the inner product. Since $(K*\rho,\nabla \rho) = -(K*\rho,\nabla \rho)$, $M^{(1)}_t = 0$ and hence the center of mass is conserved. It follows directly that (\ref{eq:ge}) does not support any solution, including unidirectional traveling solutions, in which the population undergoes a net translation. Such solutions could exist if we included additional advective terms, for instance those due to gradients in food sources. We omit such advective terms since we are primarily interested in studying the balance of aggregative and dispersive effects. Equation (\ref{eq:ge}) does \emph{not} conserve higher moments. For instance, the spread of the population is free to change under the dynamics of (\ref{eq:ge}), and indeed, will do so as the populations clumps (see Section \ref{sec:coarsen}).

\begin{figure}
\centerline{\resizebox{\textwidth}{!}{
\includegraphics{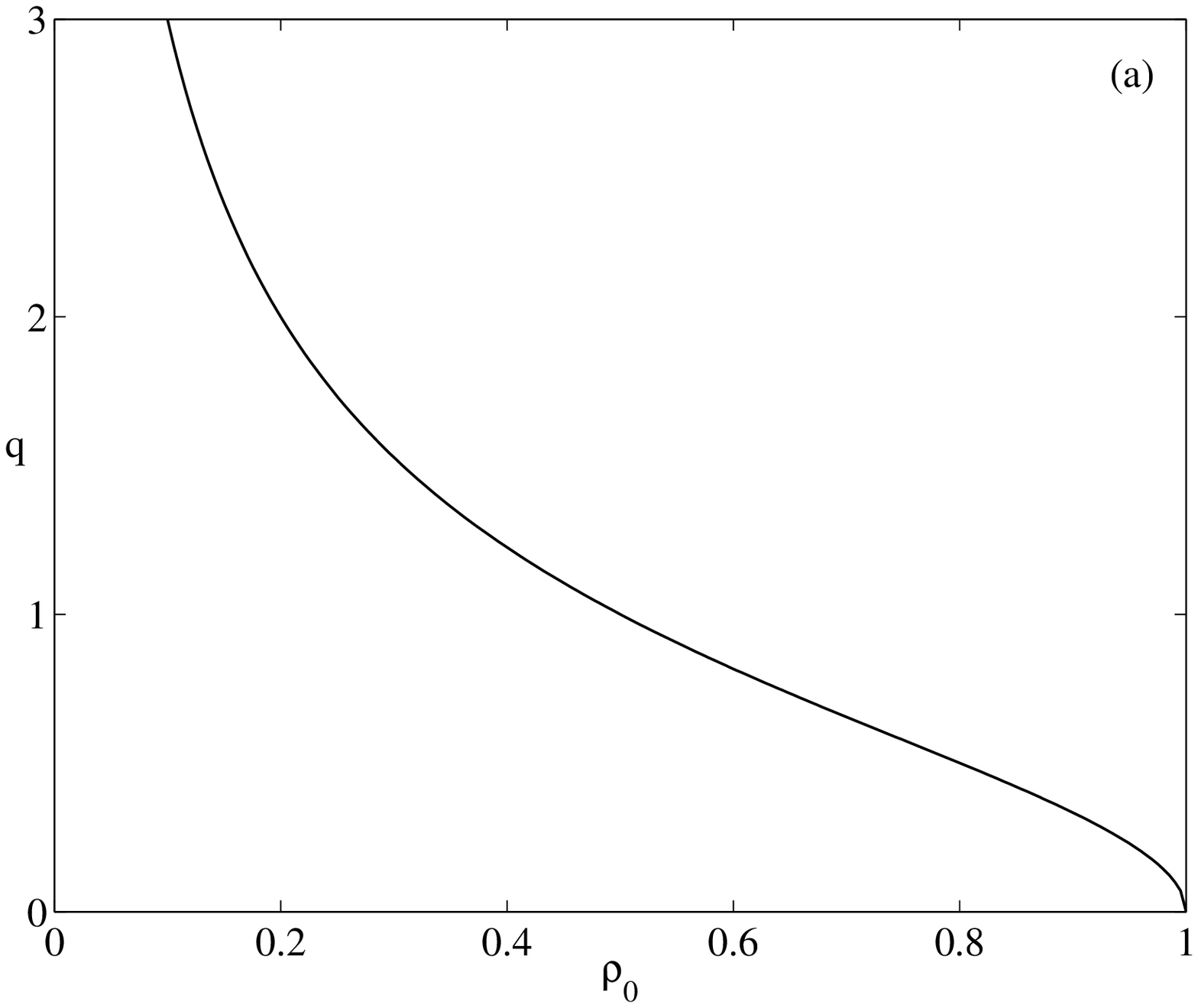} \hfill
\includegraphics{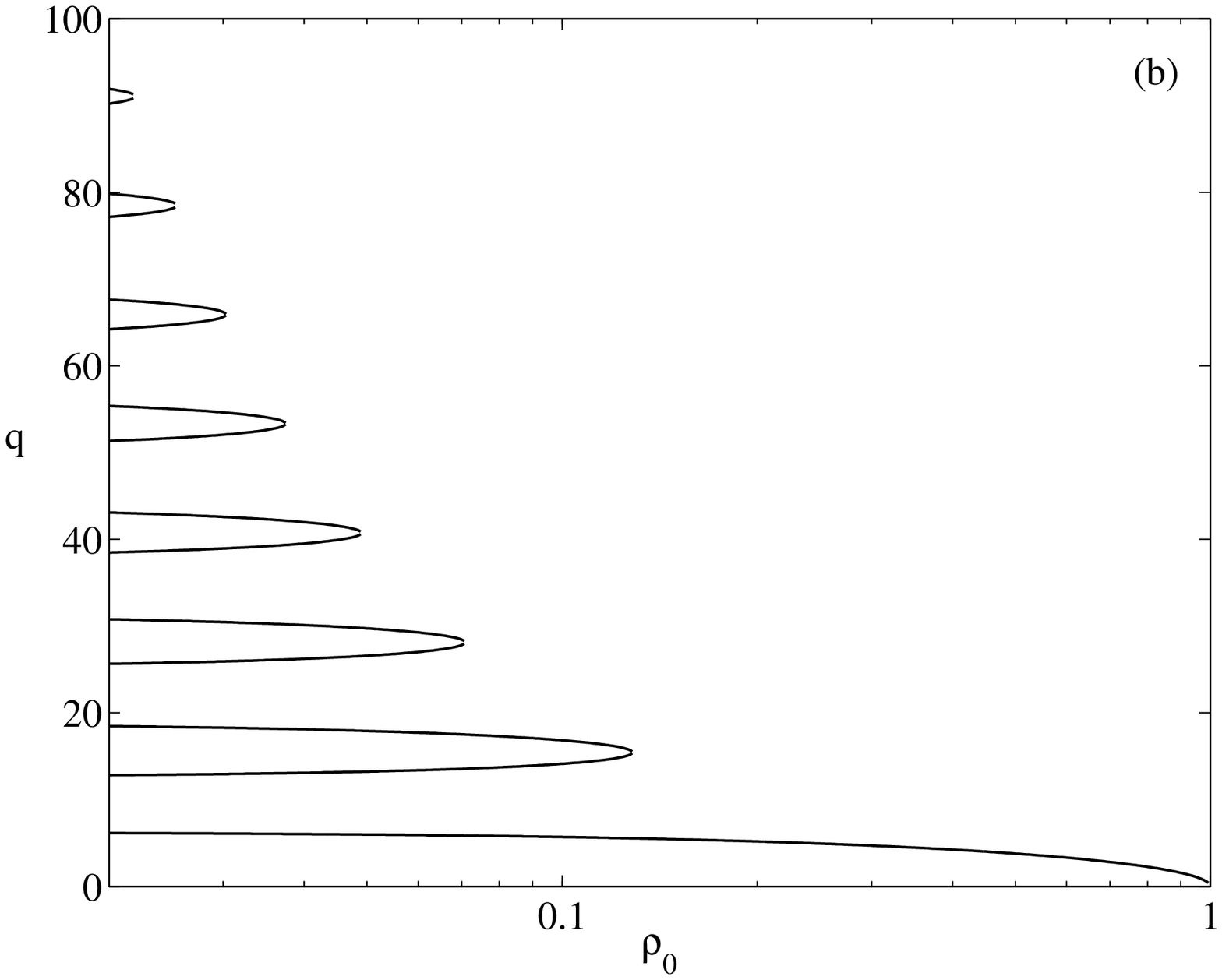}}}
\caption{(a) Neutral stability curve for (\ref{eq:ge}) calculated from the dispersion relation  $\sigma(q)$ in (\ref{eq:disprel}) with $K=K^e_1$ from (\ref{eq:k}). The region below the curve corresponds to linear instability. (b) Like (a), but with $K=K^w_1$ from (\ref{eq:k}). Since $K^w_1$ is discontinuous, $\sigma(q)$ is oscillatory and the neutral curve has multiple branches. The regions inside the curves correspond to linear instability. For both (a) and (b), $r=1$ in (\ref{eq:ge}).}
\label{fig:linear}
\end{figure}

\textbf{Linear stability.} Equation (\ref{eq:ge}) admits steady states with constant density $\rho_0 \geq 0$. To analyze their stability, we let $\rho = \rho_0 + \widehat{\rho}\exp{i\vec{q}\cdot\vec{x}+\sigma t}$ where $\vec{q}$ is a perturbation wave vector and $\sigma$ is the linear growth rate. Note that the admissible $\vec{q}$ depend on the chosen boundary conditions. For the free space problem, $\vec{q} \in \mathbb{R}^n$. For the case of periodic boundary conditions, \emph{i.e.} when $D$ is $n$-torus of size $L$, $q \in \mathbb{Z}^n \cdot 2\pi/L$. For the case of no-flux boundary conditions, $q \in \mathbb{Z}^n \cdot \pi/L$. Substituting into (\ref{eq:ge}) and retaining only the terms linear in $\widehat{\rho}$, we obtain the dispersion relation
\begin{equation}
\label{eq:disprel}
\sigma(q) = \rho_0 q^2 (\widehat{K}(q) - r \rho_0).
\end{equation}
where $q=|\vec{q}|$ and $\widehat{K}$ denotes the Fourier transform of $K$. Note that $\sigma(0) =  0$ for all $\rho_0$ due to the conservative structure of (\ref{eq:ge}) \citep{ch1993}. We determine the neutral stability curve $q(\rho_0)$ by setting $\sigma = 0$ in (\ref{eq:disprel}). Figure \ref{fig:linear} shows examples for $n=1$ spatial dimension in free space. We consider both choices of interaction kernel from (\ref{eq:k}), for which $\widehat{K}_1^e=1/(1+q^2)$ and $\widehat{K}_1^w=(1/q)\sin(q)$. For the latter case, $\widehat{K}_1^w$ is oscillatory, so there may be multiple bands of unstable $q$ which are manifest as the multi-branch curve in Figure \ref{fig:linear}b. The former case will be considered in Section \ref{sec:1dexample}.

\textbf{Energy.} Finally, we note that (\ref{eq:ge}) possesses a Lyapunov function or \emph{energy}
\begin{equation}
\label{eq:energy}
E(\rho) = \int_{D} \frac{r}{3} \rho^3 - \rho K*\rho\ d\vec{x}
\end{equation}
where the first term arises from avoidance and the second from aggregation. This energy is dissipated under the dynamics of (\ref{eq:ge}). To see this, we note
\begin{subequations}
\begin{eqnarray}
E_t & = & (\rho_t,r \rho^2) - (\rho_t,K*\rho) - (\rho,K*\rho_t) \\
& = & (\rho_t,r \rho^2) - (\rho_t,K*\rho) - (\rho_t,K*\rho) \label{eq:energy0} \\
& = & (\rho_t,r \rho^2) - 2(\rho_t,K*\rho) \\
& = & (\rho_t,r \rho^2 - 2K*\rho) \\
& = & (-\nabla \cdot [\rho K*\nabla \rho - r \rho^2 \nabla \rho], r \rho^2 -2K*\rho) \label{eq:energy1} \\
& = & (\rho K*\nabla \rho - r \rho^2 \nabla \rho, 2 r \rho \nabla \rho - 2K*\nabla \rho) \label{eq:energy2} \\
& = & -2\int_{D} \rho (|K*\nabla \rho|-r \rho |\nabla \rho|)^2\ d\vec{x} \ \leq \ 0 \label{eq:energy3}
\end{eqnarray}
\end{subequations}
where (\ref{eq:energy0}) follows from moving the convolution across the inner product, (\ref{eq:energy1}) follows from substituting (\ref{eq:ge}),  (\ref{eq:energy2}) follows from integration by parts, and the inequality in (\ref{eq:energy3}) follows since $\rho \geq 0$ because it is a density. Although Lyapunov functions appear in earlier analysis of one-dimensional swarming models \citep{a1985}, the particular energy (\ref{eq:energy}) has not been discussed before. This non-convex functional is composed of a positive avoidance term $\int_{D} \frac{1}{3} \rho^3\ d\vec{x}$ and a negative aggregation term $-\int_{D} \rho K*\rho\ d\vec{x}$ which have different nonlinear dependence on $\rho$ and different length scales. This is the hallmark of many pattern-forming systems in nature. Non-biological examples include the Cahn-Hilliard equation, which models the spinodal decomposition of binary alloys \citep{c1968,efg1989,bf1990}, droplet formation in dewetting fluid films \citep{bgw2001,ob2001,gw2003}, and self-aggregation of finite-sized particles \citep{hp2005}. These models are well-known to exhibit coarsening dynamics, in which small localized clumps form and merge into larger clumps over time. We will demonstrate that~(\ref{eq:ge}) exhibits similar behavior.

\section{Case study in one spatial dimension}
\label{sec:1dexample}

We now consider in detail (\ref{eq:ge}) posed in one spatial dimension, \emph{i.e.} $n=1$. Our primary goal is to determine the possible steady states for different population sizes, and thus we conduct a bifurcation study with the population size $M$ as the bifurcation parameter. We derive analytical expressions for weakly nonlinear solutions, and use phase plane, energy, and numerical methods to investigate the more nonlinear clumped solutions. We take $K=K^e_1$, the exponential kernel in (\ref{eq:kexp}), because of the biologically reasonable constant hazard function assumption that leads to it. For mathematical convenience, we begin by considering a periodic domain, which requires that $K^e_1$ be modified so that it is likewise periodic. A simple way to do this is to note that in free space, the operator $K^e_1*$ is the inverse of the operator $I-\partial_x^2$. Thus, a natural modification is to choose a periodic kernel which is precisely the inverse of $I-\partial_x^2$ on a periodic box. This way, the free-space and periodic kernels have Fourier transforms which coincide (or more precisely, the continuous transform of the free-space kernel interpolates the discrete transform of the periodic one). In Section \ref{sec:nonlin}, in order to approximate (\ref{eq:ge}) in free space, we take the limit of large domain lengths $L$. Furthermore, we note that to study the case of no-flux boundary conditions, one may simply construct an even extension of the problem, which then has periodic boundary conditions.

\subsection{Weakly nonlinear analysis}
\label{sec:wna}

We pose (\ref{eq:ge}) on a periodic box $D = [0,L]$ of length $L=2\pi/q$ and consider the instability of the constant density steady state as the total population size $M$ is decreased through the critical value
\begin{equation}
M^c =  L \rho_c = \frac{2\pi \rho_c}{q} = \frac{2 \pi \widehat{K}(q)}{qr}
\end{equation}
where $\rho_c$ is the critical constant density for the given box size, as determined by the dispersion relation (\ref{eq:disprel}). We seek weakly nonlinear solutions of (\ref{eq:ge}) via a multiple-scales perturbation expansion, letting
\begin{subequations}
\begin{eqnarray}
\partial_t & \rightarrow & \partial_t + \epsilon^2 \partial_T\\
M & = & M_c + \epsilon^2 m \\ 
\rho(x,t) & = & \rho_0 + \epsilon \rho_1(x,T) + \epsilon^2 \rho_2(x,T) + \epsilon^3 \rho_3(x,T) \ldots\ .
\end{eqnarray}
\end{subequations}
Here, $\epsilon \ll 1$ is a small bookkeeping parameter, $T$ is a slow time scale, and $m$ is the deviation from the critical population size. The $\rho_i$ do not evolve on the fast time scale $t$, and so the $t$ dependence is henceforth eliminated from our calculation. The perturbation expansion allows us to compute a solution to the nonlinear homogeneous problem (\ref{eq:ge}) in terms of the expansion functions $\rho_0(x,T)$, $\rho_1(x,T)$ and so on, where these functions satisfy linear nonhomogeneous equations. Also,
\begin{equation}
\rho_0 = M/L
\end{equation}
by definition, and
\begin{equation}
\rho_1 = z(T)\exp{iqx}+c.c.
\end{equation}
where $c.c.$ stands for complex conjugate and $z$ is the slowly varying amplitude of the solution, for which we will derive an amplitude equation.

At $\order(\epsilon)$, (\ref{eq:ge}) becomes the linear problem
\begin{equation}
\rho_c K * \rho_{1,xx} - r \rho_c^2 \rho_{1,xx} = 0
\end{equation}
which simply recovers the dispersion relation (\ref{eq:disprel}). At $\order(\epsilon^2)$, we obtain
\begin{eqnarray}
\rho_c K * \rho_{2,xx} - r \rho_c^2 \rho_{2,xx} & = & -\rho_{1,x} K*\rho_{1,x} - \rho_1 K*\rho_{1,xx} \\
& & \mbox{} + 2r\rho_c (\rho_{1,x})^2 + 2r\rho_c \rho_1 \rho_{1,xx} \nonumber
\end{eqnarray}
which has the particular solution
\begin{equation}
\rho_2 = -\frac{r}{2\widehat{K}(q)-2 \widehat{K}(2q)}z^2(T) \exp{2iqx}+c.c.\ .
\end{equation}
At $\order(\epsilon^3)$, we obtain
\begin{eqnarray}
\rho_c K * \rho_{3,xx} - r \rho_c^2 \rho_{3,xx} & = & \mbox{}-\rho_{1,T} - \rho_{1,x} K * \rho_{2,x} - \rho_{2,x} K * \rho_{1,x} \label{eq:cubicorder} \\
& & \mbox{} - \rho_1 K * \rho_{2,xx} - (m/L)K * \rho_{1,xx} \nonumber \\
& & \mbox{} - \rho_2 K * \rho_{1,xx} + 4r\rho_c\rho_{1,x}\rho_{2,x} + 2r\rho_1(\rho_{1,x})^2 \nonumber \\
& & \mbox{} + 2 r \rho_c \rho_1 \rho_{2,xx} + r \rho_1^2 \rho_{1,xx} + 2r \rho_c \rho_2 \rho_{1,xx} \nonumber \\
& & \mbox{} + 2r(m/L)\rho_c \rho_{1,xx}. \nonumber
\end{eqnarray}
The right hand side of (\ref{eq:cubicorder}) contains terms with spatial dependence $\exp{iqx}$ which lie in the solution space of the left hand side. To guarantee that $\rho$ is periodic, we must eliminate these secular terms. Enforcing this solvability condition leads to the amplitude equation
\begin{equation}
\frac{\d z}{\d T} = \lambda z + a |z|^2z, \quad
\lambda = -m \widehat{K}(q)q^2/L	, \quad a = \frac{r q^2 \widehat{K}(q)}{2\widehat{K}(q) - 2\widehat{K}(2q)}.
\label{eq:ampeq}
\end{equation}
This result is valid for any $K$ satisfying the assumptions made in Section \ref{sec:model}.

Equation (\ref{eq:ampeq}) describes a pitchfork bifurcation \citep{c1991a} from the constant density state to a spatially-periodic pattern with steady-state amplitude determined by $|z|^2=-\lambda m/a$. For $K=K^e_1$ from (\ref{eq:kexp}), $a>0$ and so the patterned state bifurcates subcritically, and hence unstably. In fact, this is the case for any $K$ and $q$ satisfying $\widehat{K}(q) > \widehat{K}(2q)$.

In Section \ref{sec:bifurcation} we will compare the result in (\ref{eq:ampeq}) with numerically obtained results. The quantity we will study is the amplitude $A = \max_{D} \rho - \min_{D} \rho$. Keeping the first two terms in our perturbation expansion, \emph{i.e.} $\rho = \rho_0 + \epsilon \rho_1$, and substituting (\ref{eq:ampeq}), we arrive at the weakly nonlinear prediction
\begin{equation}
A = \frac{4}{r}\sqrt{\frac{2}{L}\left( Mr-\widehat{K}(q)L\right) \left(\widehat{K}(q) - \widehat{K}(2q)\right) }.
\end{equation}
It follows that the amplitude of the periodic pattern is inversely proportional to $r$, or directly proportional to the ratio of typical attractive speed to typical repulsive speed.

\subsection{Phase plane analysis for steady states}
\label{sec:pp}

\begin{figure}
\centerline{\resizebox{\textwidth}{!}{
\includegraphics{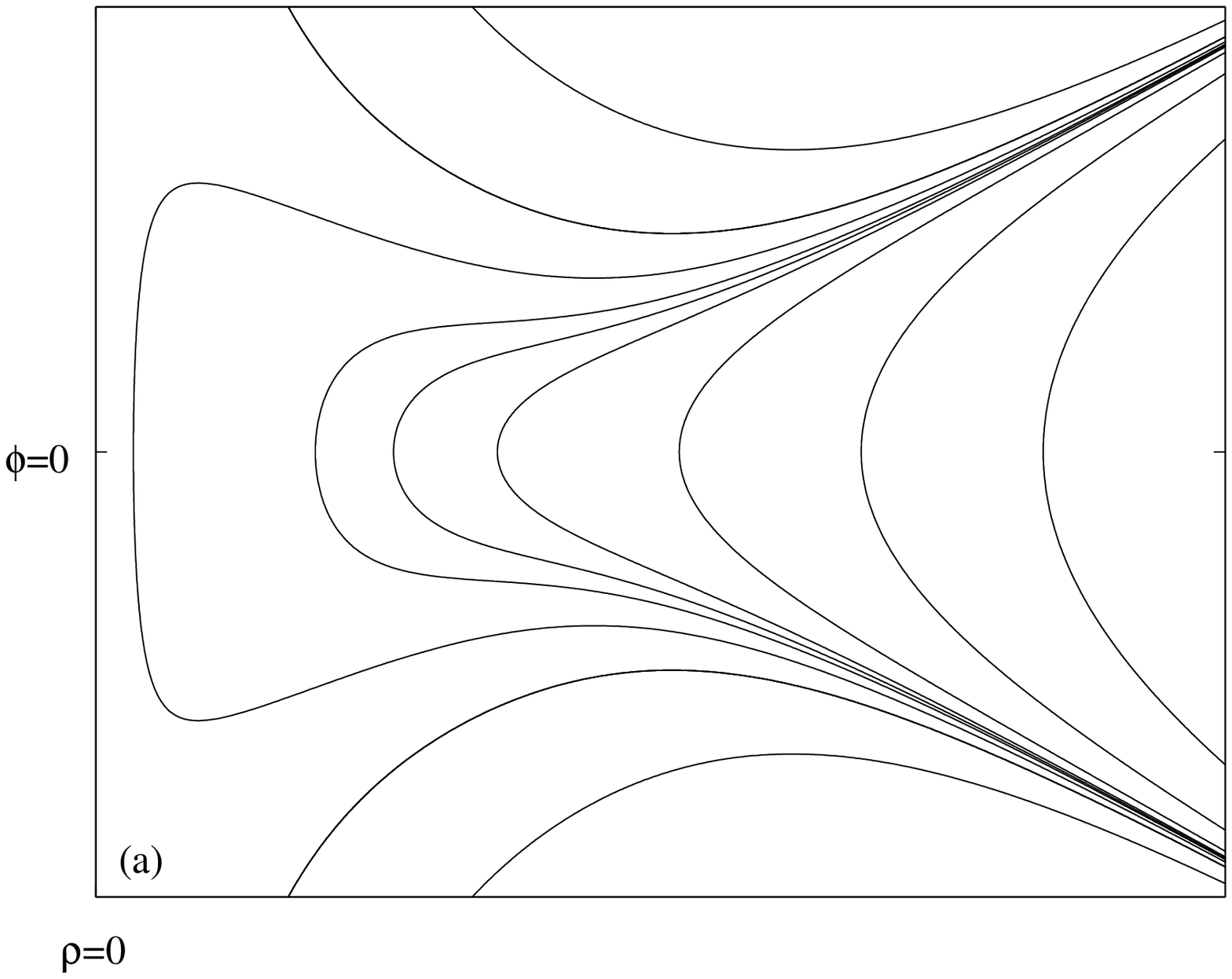} \hfill
\includegraphics{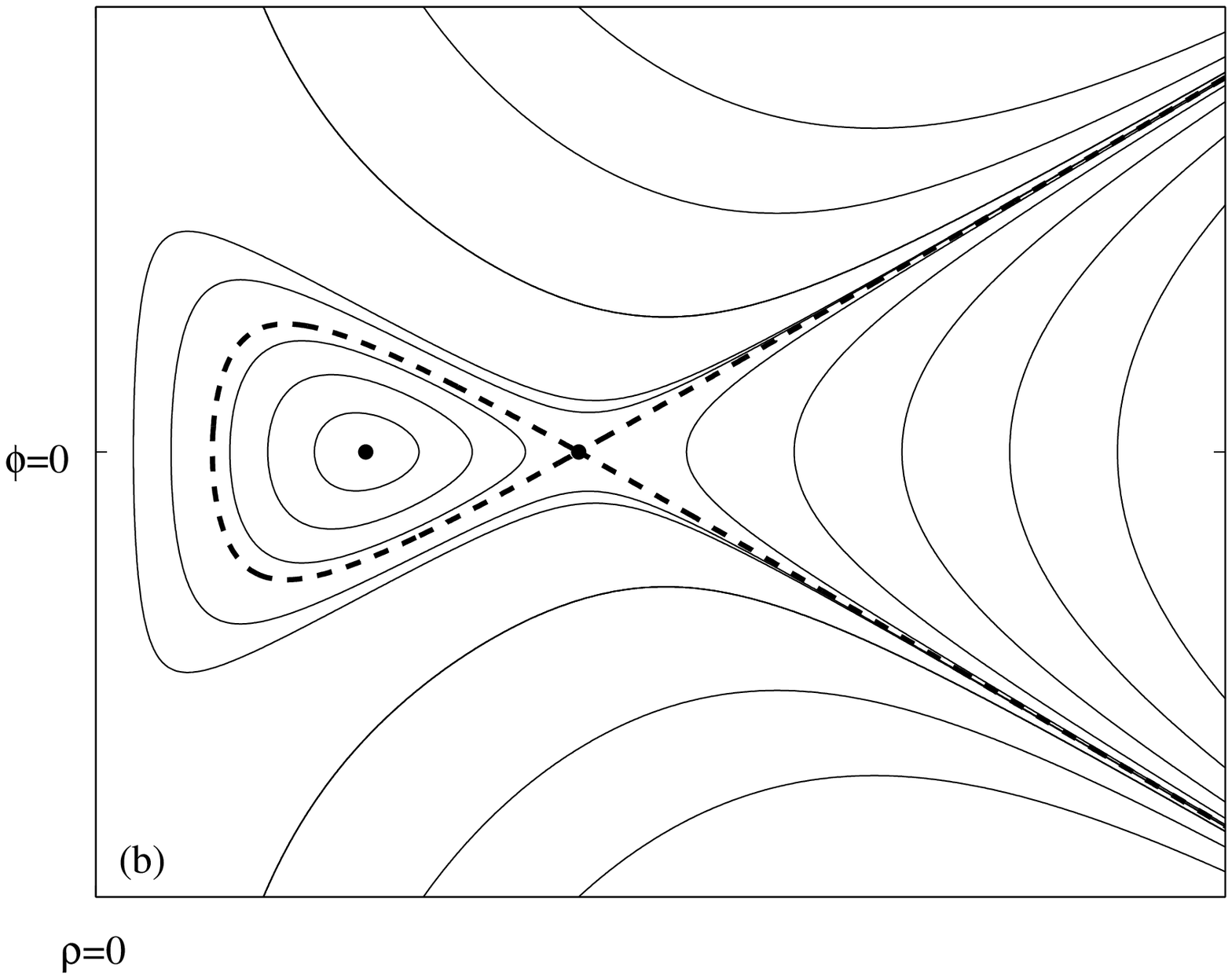}}}
\centerline{\resizebox{\textwidth}{!}{
\includegraphics{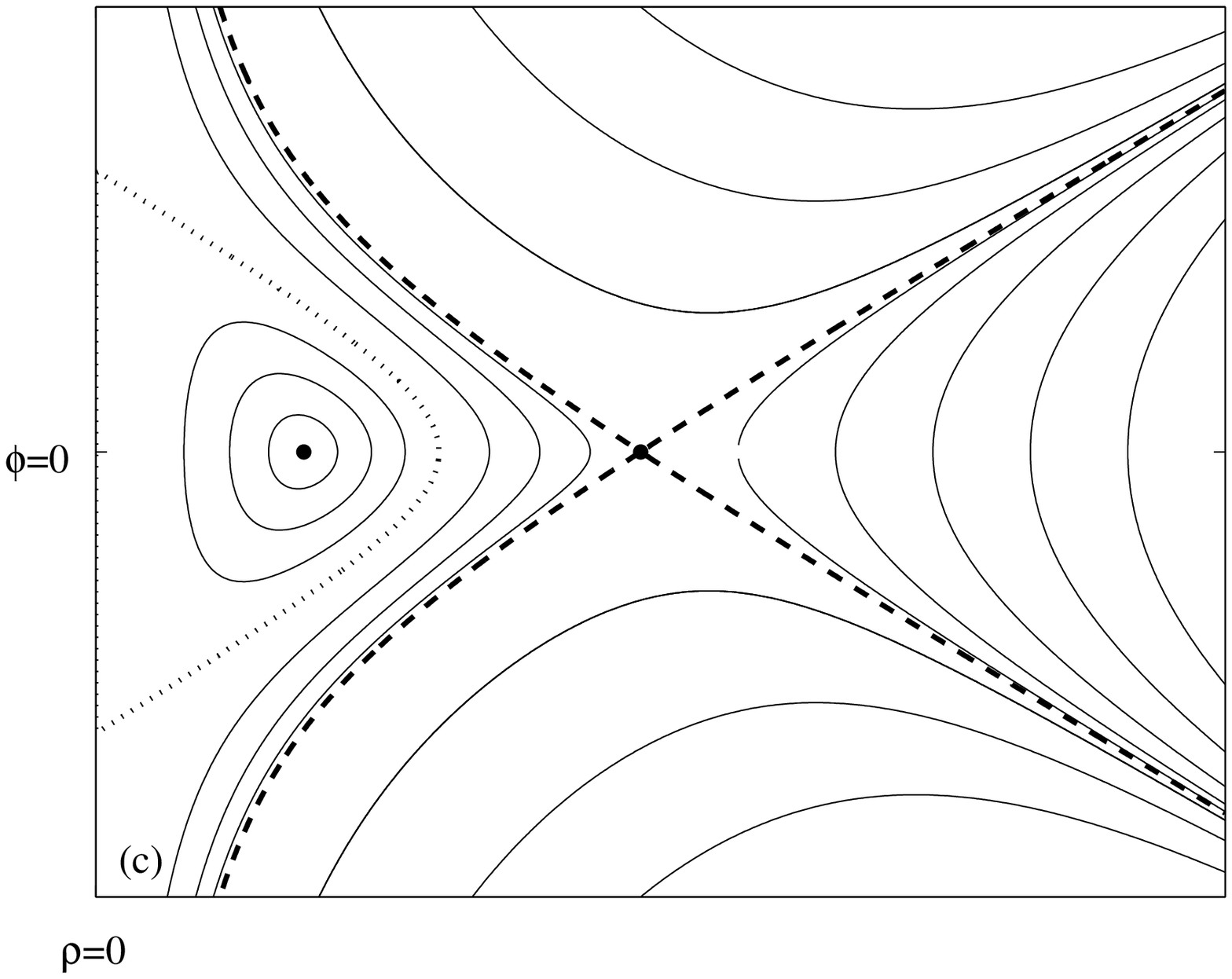} \hfill
\includegraphics{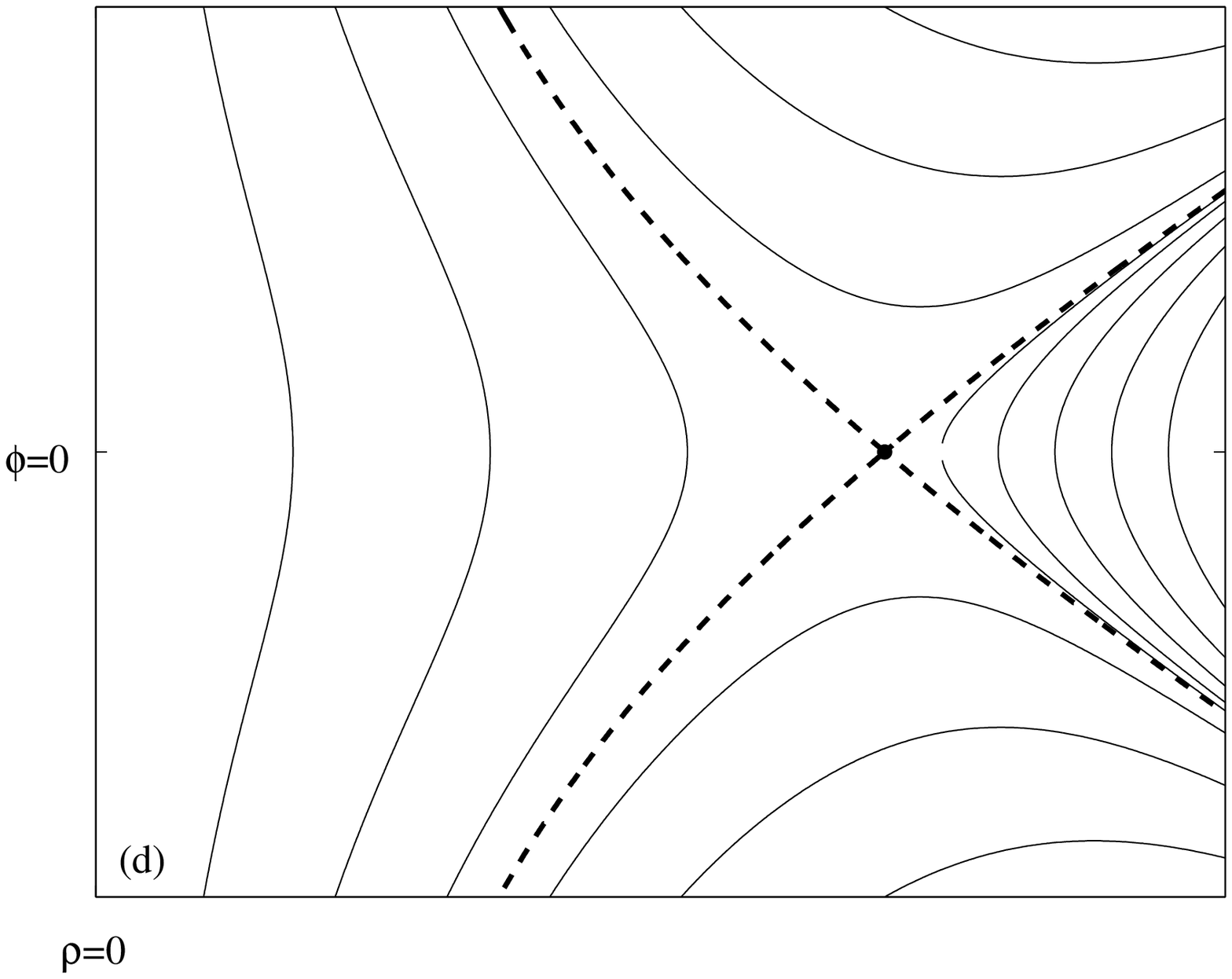}}}
\caption{(a) - (d): phase portraits for (\ref{eq:pp}) corresponding to different values of $C$ for the cases cases I,III,V,VII described in detail in the text. The phase plane problem describes the steady states of (\ref{eq:ge}). The stable and unstable manifolds of the fixed point $F^+$ are indicated by the broken line, and the critical touchdown solution $T$ is indicated by the dotted line.}
\label{fig:pp}
\end{figure}

We set the flux in (\ref{eq:ge}) equal to $0$ and divide by $\rho$ (discarding the trivial solution) to obtain the steady state problem $K^e_1*\rho_x - r \rho \rho_x = 0$.
Inverting $K^e_1$ yields the \emph{local} steady state equation
\begin{equation}
\rho_x = r \rho \rho_x - r (\rho \rho_x)_{xx} \label{eq:ss}
\end{equation}
which may be integrated once to obtain $\rho = r \rho^2/2 - r (\rho \rho_x)_x + C/2$, where $C/2$ is the constant of integration. This equation has an exact solution in terms of elliptic integrals, but it is difficult to analyze. Instead, we study the equivalent phase-plane problem
\begin{gather}
\dot{\rho} = \phi, \quad \dot{\phi} = -\frac{1}{r} + \frac{\rho}{2} - \frac{\phi^2}{\rho} + \frac{C}{2r\rho} \label{eq:pp}\\
\mathcal{E}(\rho,\phi,C) = \frac{\rho^3}{3} - \frac{r \rho^4}{8} + \frac{r \rho^2\phi^2}{2} - \frac{C \rho^2}{4} \label{eq:ppenergy}
\end{gather}
where $\phi=\rho_x$, the dot represents $d/dx$, and $\mathcal{E}$ is the ``energy'' (different from the energy functional in equation \ref{eq:energy}) which is constant on trajectories of the steady state problem (\ref{eq:pp}). We classify the solutions as the parameter $C$ is varied. Cases I, III, V, and VII below are depicted in Figure~\ref{fig:pp}.

\emph{Case I: $C > 1/r$.} There are no finite attractors. All trajectories are unbounded.

\emph{Case II: $C=1/r$.} A fixed point is born at $(\rho,\phi)=(1/r,0)$.

\emph{Case III: $8/(9r) < C < 1/r$.} There are two fixed points $F^{\pm}$ corresponding to constant density populations, located at $(\rho,\phi)=(\rho^{\pm},0)$, $\rho^{\pm} = (1 \pm \sqrt{1-rC})/r$. $F^+$ is a saddle whose stable/unstable manifold $W^{s,u}$ forms a homoclinic loop which is bounded away from the $\phi$ axis. $F^-$ is a nonlinear center, and lies inside $W^{s,u}$. Between $F^-$ and $W^{s,u}$ lies a continuous family of periodic orbits which correspond to smooth, patterned populations.

\emph{Case IV: $C = 8/(9r)$.} $W^{s,u}$ collides with the origin.

\emph{Case V: $0 < C < 8/(9r)$.} The homoclinic loop is destroyed, \emph{i.e.} $W^{s,u}$ no longer self-intersects. Periodic orbits are destroyed as they collide with the origin. There exists a critical orbit corresponding to a population with peak density  $\rho_{max}=(4 - \sqrt{16-18rC})/3$. This \emph{touchdown solution} just touches down to $\rho = 0$ and does so with finite slope $\phi^T = \pm \sqrt{C/(2r)}$. The region bounded by the touchdown trajectory, the $\phi$-axis, and $W^{s,u}$ contains a continuous family of trajectories which approach $(\rho,\phi)=(0,\pm\infty)$. For $\rho\ll1$ and $|\phi|\gg 1$, (\ref{eq:pp}) is approximately $\rho\ddot{\rho} + (\dot{\rho})^2=0$, which has positive nontrivial solutions of the form $\sqrt{ax+b}$. Thus, solutions in $S$ reach $\rho=0$ for finite $x$ and correspond to compactly supported swarm-like populations which touch down with \emph{infinite} slope.

\emph{Case VI: $C=0$.} $F^-$ collides with the origin.

\emph{Case VII: $C<0$.} The periodic orbits are destroyed. The only bounded solutions are the compactly supported ones, lying between $W^{s,u}$ and the $\phi$-axis.

In summary, for  $8/(9r) < C < 1/r$, only trivial (constant density) and periodic solutions are possible. For $0 < C < 8/(9r)$, trivial, periodic, and clumped solutions are possible. For $C<0$, only trivial and clumped solutions are possible. Note that the solutions of (\ref{eq:pp}) actually form a two parameter family. One parameter is the mathematical constant of integration $C$, which selects a particular phase portrait. The other parameter may be taken to be the ``initial condition'' in that phase portrait, which selects a particular trajectory. More biologically relevant parameters include the total population size $M$, the peak density $\rho_{max}$, and the period or support of the steady solution. In fact, fixing any two of these quantities determines the third. Unfortunately, the relationship between these biological quantities and the mathematical ones we have used for our phase plane study cannot be determined by the qualitative analysis we have carried out in this section. In order to understand the role of the biologically relevant quantities, we perform a numerical investigation in the next section.

\subsection{Numerical solution and bifurcation diagram}
\label{sec:bifurcation}

\begin{figure}
\centerline{\resizebox{\textwidth}{!}{\includegraphics{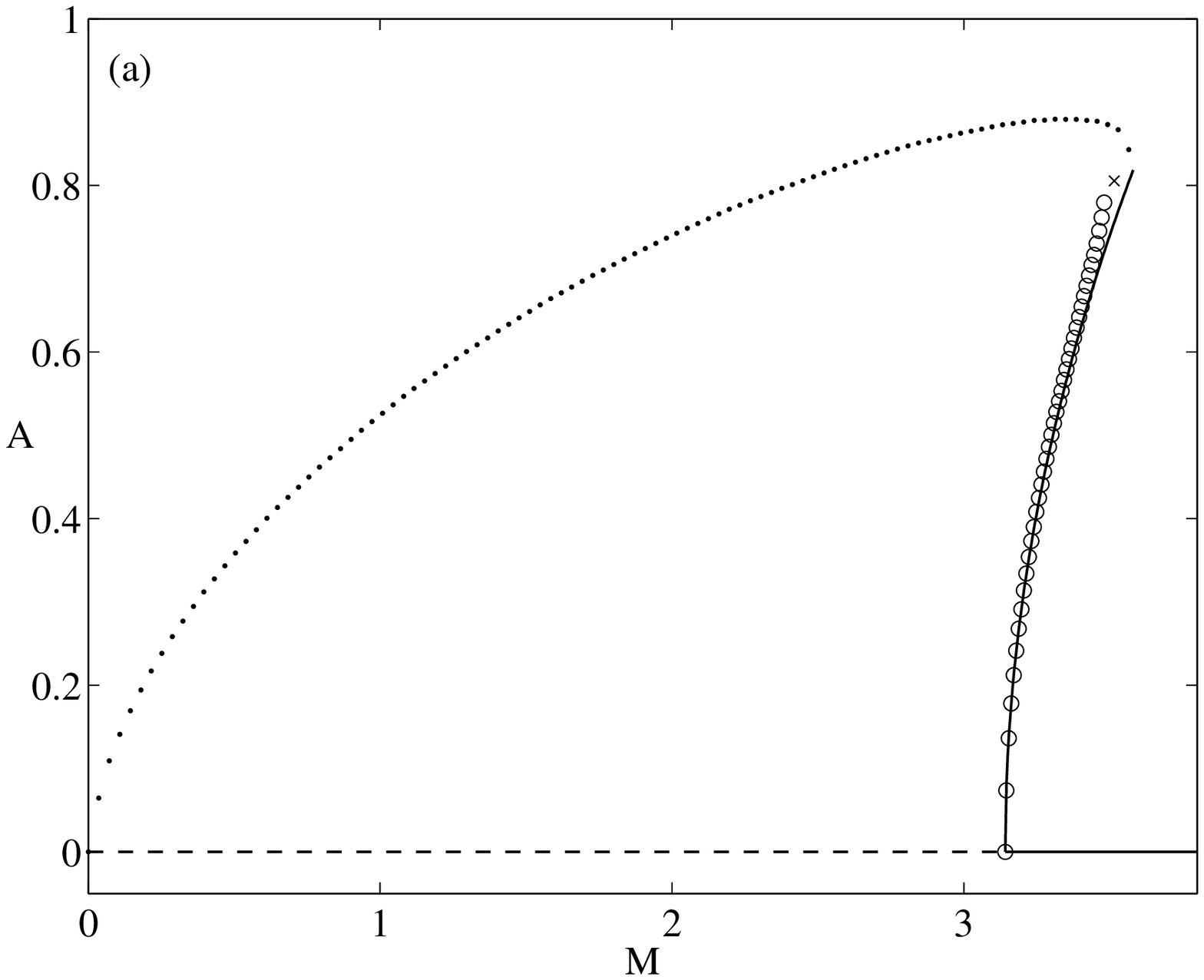} \hspace{0.1in} \includegraphics{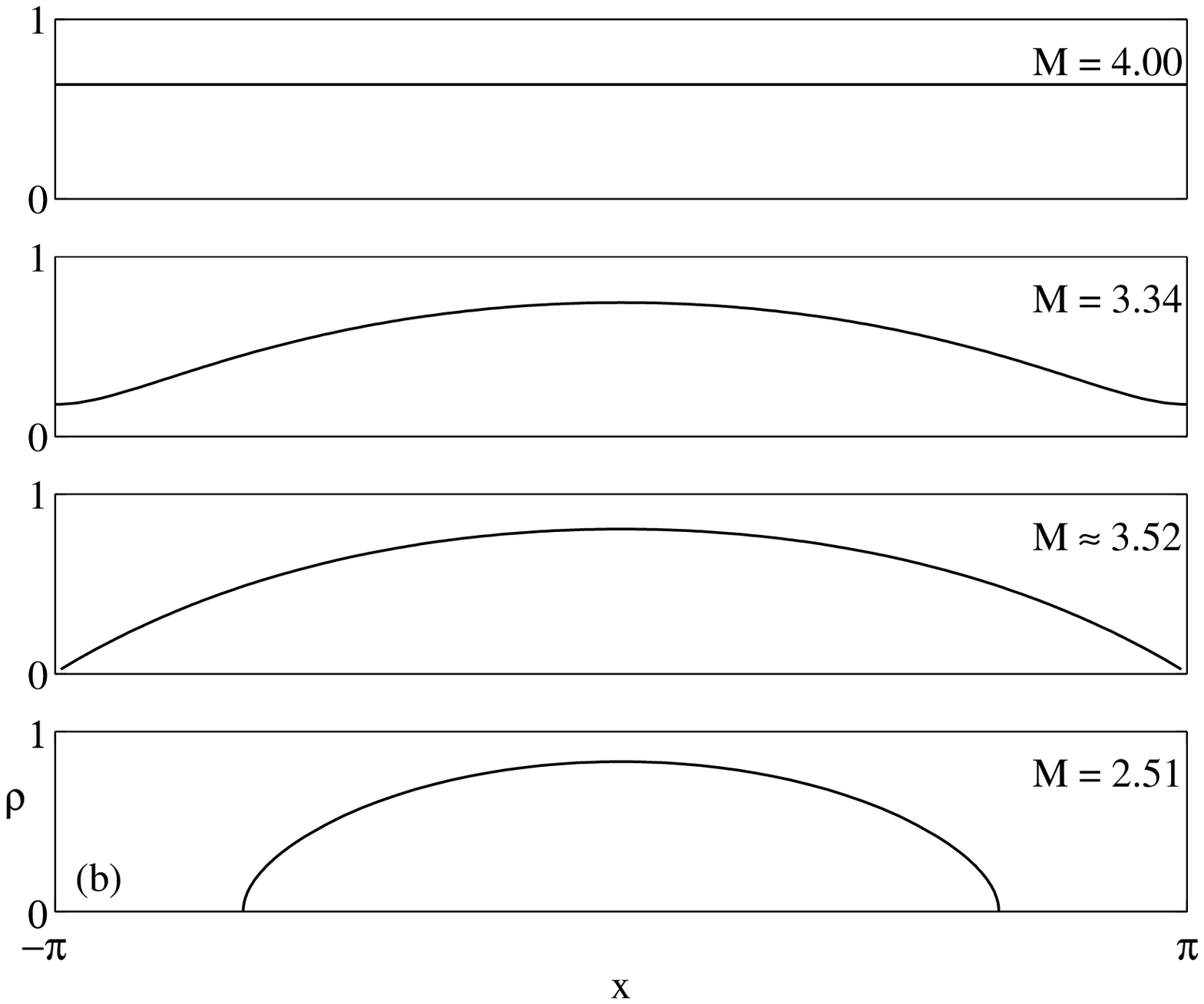}}}
\caption{(a) Bifurcation diagram for (\ref{eq:ge}) with $K=K^e$ from (\ref{eq:kexp}) depicting solution amplitude $A$ versus the total population size $M$. The size of the periodic box is $L=2\pi$, and we take $r=1$. The different solution types are stable constant density (horizontal sold line), unstable constant density (dashed line), periodic (open circles), touchdown periodic (x), and compactly supported (dots). The solid curve is the bifurcation curve predicted by (\ref{eq:ampeq}), the result of a weakly nonlinear analysis. (b) Steady state solutions for different values of M corresponding to different points along the bifurcation curve in (a).}
\label{fig:bifurcation}
\end{figure}

We fix $L$ and construct the  bifurcation diagram describing the steady state solutions of~(\ref{eq:ge}). An example  for $L=2\pi$ with $r=1$ is given in Figure \ref{fig:bifurcation}a, which shows the solution amplitude $A = \max_{D} \rho - \min_{D} \rho$ as the population size $M$ is varied.  Solution profiles corresponding to different values of $M$ along the bifurcation curve are sown in Figure \ref{fig:bifurcation}b. The bifurcation diagram and solution profiles are obtained in the following way. 

For the periodic and touchdown solutions, we compute trajectories of  (\ref{eq:pp}) numerically using the Runge-Kutta (4,5) method, locate the trajectory with period $L$, and measure $M$ (using the trapezoidal rule) and $A$. This process is performed different values of $C$. We perform a similar procedure for the clump solutions. However, for fixed $C$, any clump solution with the correct mass and with support that fits inside $D$ is allowed. Thus, for fixed $M$, there is not a unique clump, but rather a continuous one-parameter family, of which each member has a different length and peak density. There is a selection mechanism which prefers one member of this family for each $M$ (in particular, the member represented in Figure \ref{fig:bifurcation}a). We discuss this mechanism in Section~\ref{sec:nonlin}.

Figure \ref{fig:bifurcation}a reveals that the subcritical branch of periodic solutions connects to the clump solutions at the touchdown solution. The subcritical branch of clump solutions turns around, and eventually terminates at the empty population solution $M=0$. From the bifurcation diagram we see that as $M$ is decreased through the bifurcation point and the trivial state destabilizes, the transition will involve a jump to a large-amplitude swarm. One interesting biological prediction which follows is that of hysteretic transitions for organisms in a confined environment (in which case the appropriate no-flux boundary conditions may be related to the periodic ones we use here; see previous discussion). Large, homogeneous populations will not form clumps. If sufficiently many organisms are removed, so as to bring the population size below the critical one, clumping will occur. Then, if organisms are added back in, the population may remain clumped even for population sizes above the critical one (so long as not too many are added). 

\begin{figure}
\centerline{\resizebox{3in}{!}{\includegraphics{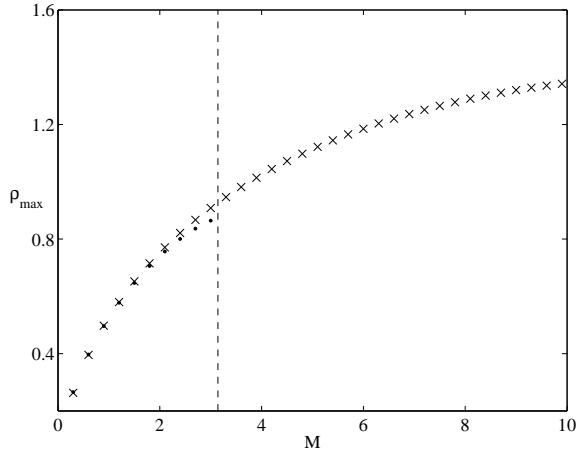}}}
\caption{Peak density $\rho_{max}$ versus population size $M$ of clump solutions as computed on periodic domains of length $L=2\pi$ (dots) and $4\pi$ (x's), both with $r=1$. The two data sets coincide for small $M$, but diverge for larger $M$ when ``wraparound'' effects due to periodicity come into play. The $L=2\pi$ branch terminates for smaller $M$ since compactly supported solutions with larger $M$ do not fit in the domain. The dotted vertical line indicates the critical population size for $L=2\pi$, namely $M_c = \pi$ (\emph{cf.} Figure \ref{fig:bifurcation}).}
\label{fig:finitesizeeffect}
\end{figure}

The particular form of the connection to the connection of the periodic branch to the branch of clumps is an artifact of having a finite periodic domain. As discussed in Section \ref{sec:model}, for the periodic problem, the kernel $K^e_1$ must be made periodic, and hence depends on the domain length $L$. Therefore, steady state solutions of identical mass, but on different domains, will not necessarily be identical. Figure \ref{fig:finitesizeeffect} demonstrates this effect. We plot the maximum density $\rho_{max}$ versus population size $M$ of compactly supported solutions as computed on periodic domains of length $L=2\pi$ and $4\pi$. The two results agree for small $M$, but diverge for larger $M$. One intuitive way of understanding this phenomenon is to imagine that for clumps with sufficiently large support, the left and right sides of the clump may interact strongly due to the nonlocal aggregative term in (\ref{eq:ge}) and the periodic boundary conditions.

\subsection{Coarsening}
\label{sec:coarsen}

It is of interest to investigate how the steady state clumps just discussed are approached. Numerical simulations reveal that our model displays ``coarsening'' dynamics. We use a fully implicit numerical scheme. Derivatives are calculated using central finite differences, and the convolution operator $K_1^e*$ is computed as $(I-\partial_x^2)^{-1}$, as discussed previously. The nonlinear problem is solved via Newton iteration.

\begin{figure}
\centerline{\resizebox{\textwidth}{!}{\includegraphics{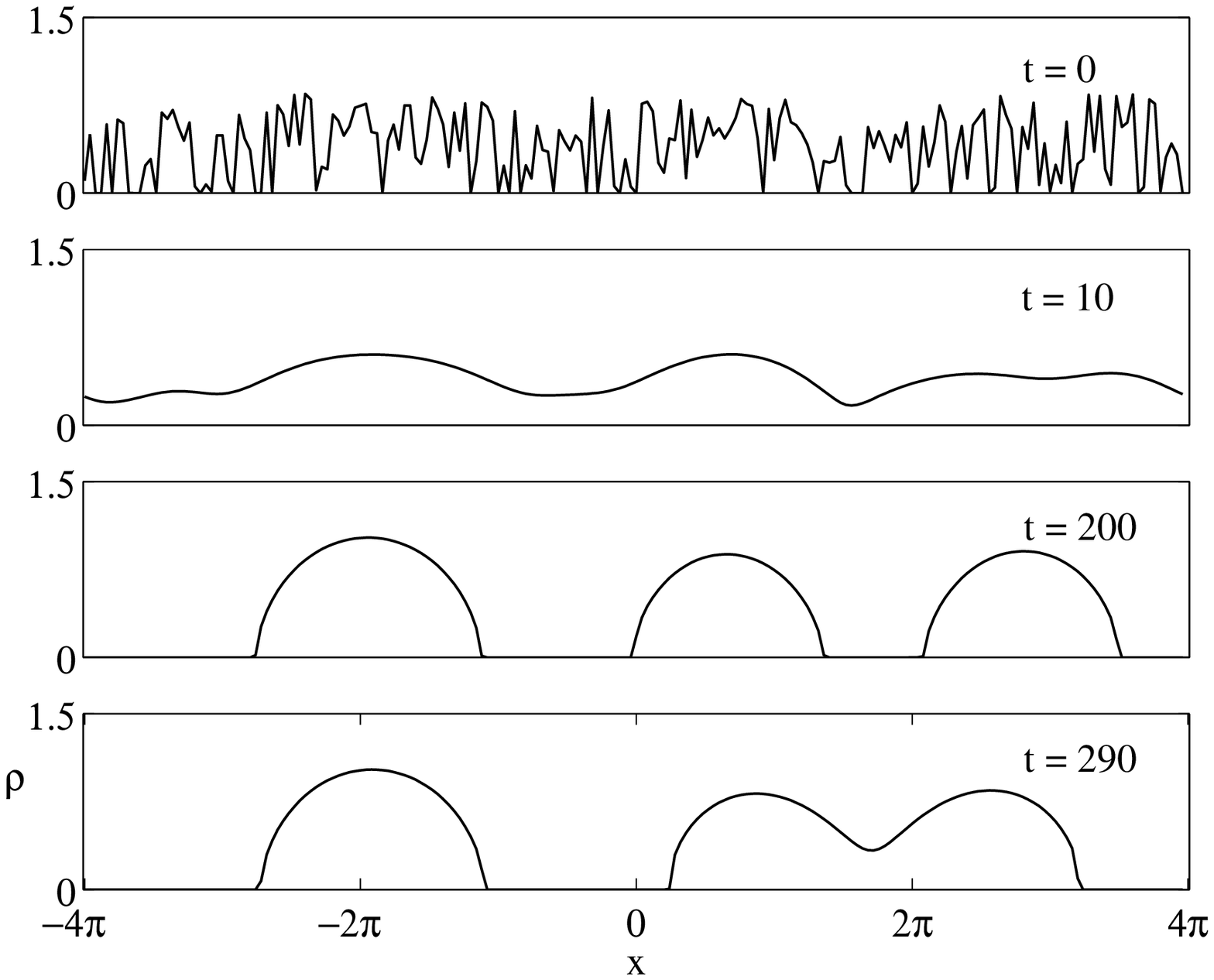} \hspace{0.1in} \includegraphics{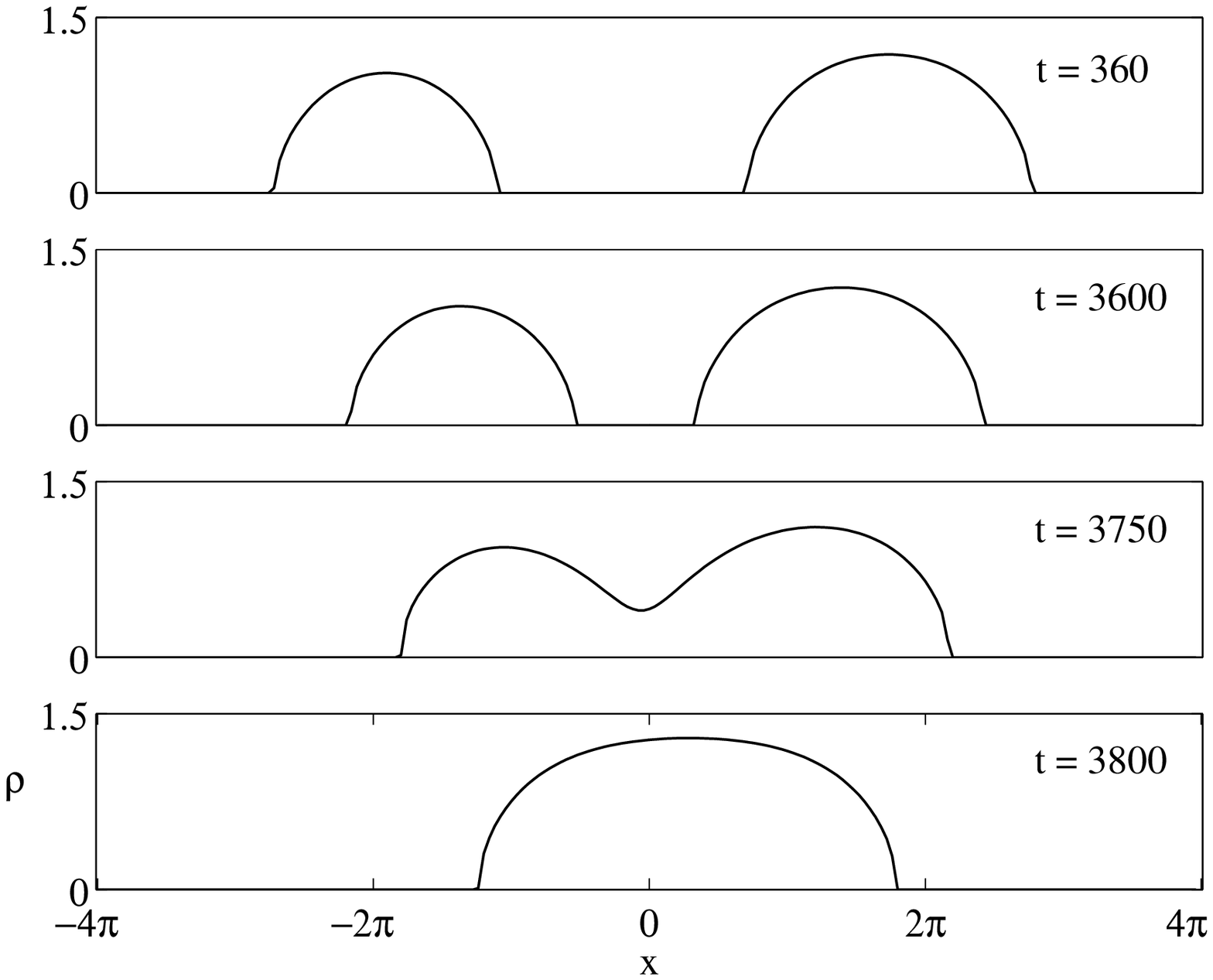}}}
\caption{Snapshots showing the coarsening of an initially randomly-distributed population with $M=10$ on a domain of length $L=8\pi$. Here, $r=1$ in (\ref{eq:ge}).}
\label{fig:coarsen}
\end{figure}

As an example, we perform simulations on a periodic box of length $L=8\pi$ and choose a random initial condition with total integral $M=10$. Figure \ref{fig:coarsen} shows a series of snapshots of the solution as it is evolved according to (\ref{eq:ge}). The irregular initial profile is rapidly smoothed. Then a more slow transition occurs, in which the smooth profile breaks up, in this case into three distinct clumps of organisms. However, due to the long-range social attraction modeled in $K$, each clump can feel the presence of the others. The two rigthmost clumps move slowly towards each other and merge into a larger group. Then, this large group and the remaining group are attracted together, though the motion occurs over an even slower time scale. Eventually, the two groups merge into one large group, and steady state is reached.

The coarsening dynamics are of special interest to us given the widely held view that ``social behaviors that on short time scales lead to the formation of social groups$\dots$at the largest time and space scales [have] profound consequences for ecosystem dynamics and for the evolution of behavioral, morphological, and life history traits'' \citep{ogk2001}. The vast majority of models for splitting and joining of social groups on long time and space scales are stochastic, and involve quantities such as the probability of a group splitting into k smaller groups, which may be extremely difficult to measure experimentally. Coarsening dynamics are a natural context in which to study the splitting and merging of groups. Moreover, this formulation has the advantage that results (such as the coarsening rate) could be explicitly tied to the rules for movement (for example, the properties of the interaction kernel $K$, or the value of the speed ratio $r$).

\subsection{Nonlinear selection and large-domain, large-population asymptotics}
\label{sec:nonlin}

\begin{figure}
\centerline{\resizebox{\textwidth}{!}{\includegraphics{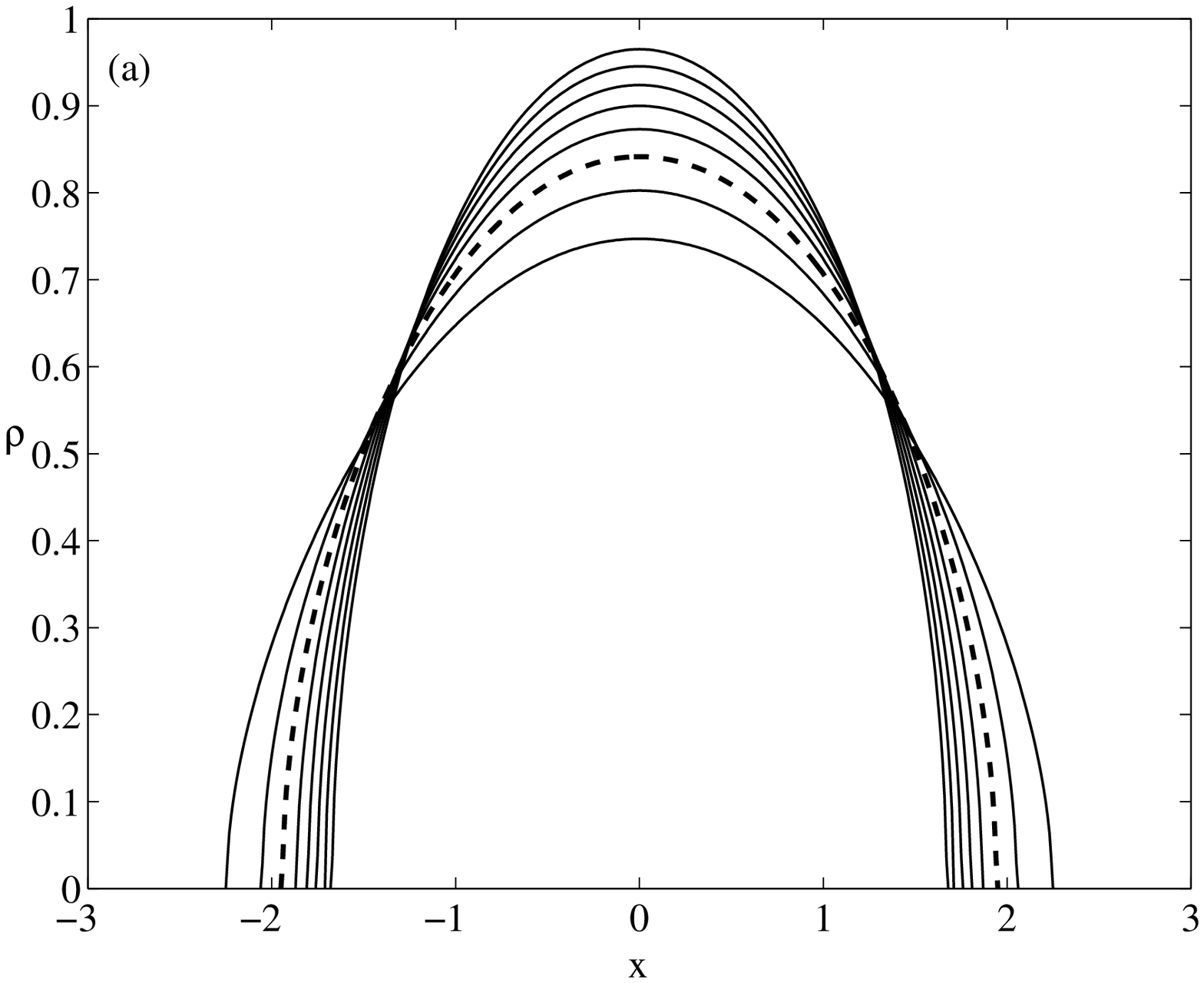} \hspace{0.1in} \includegraphics{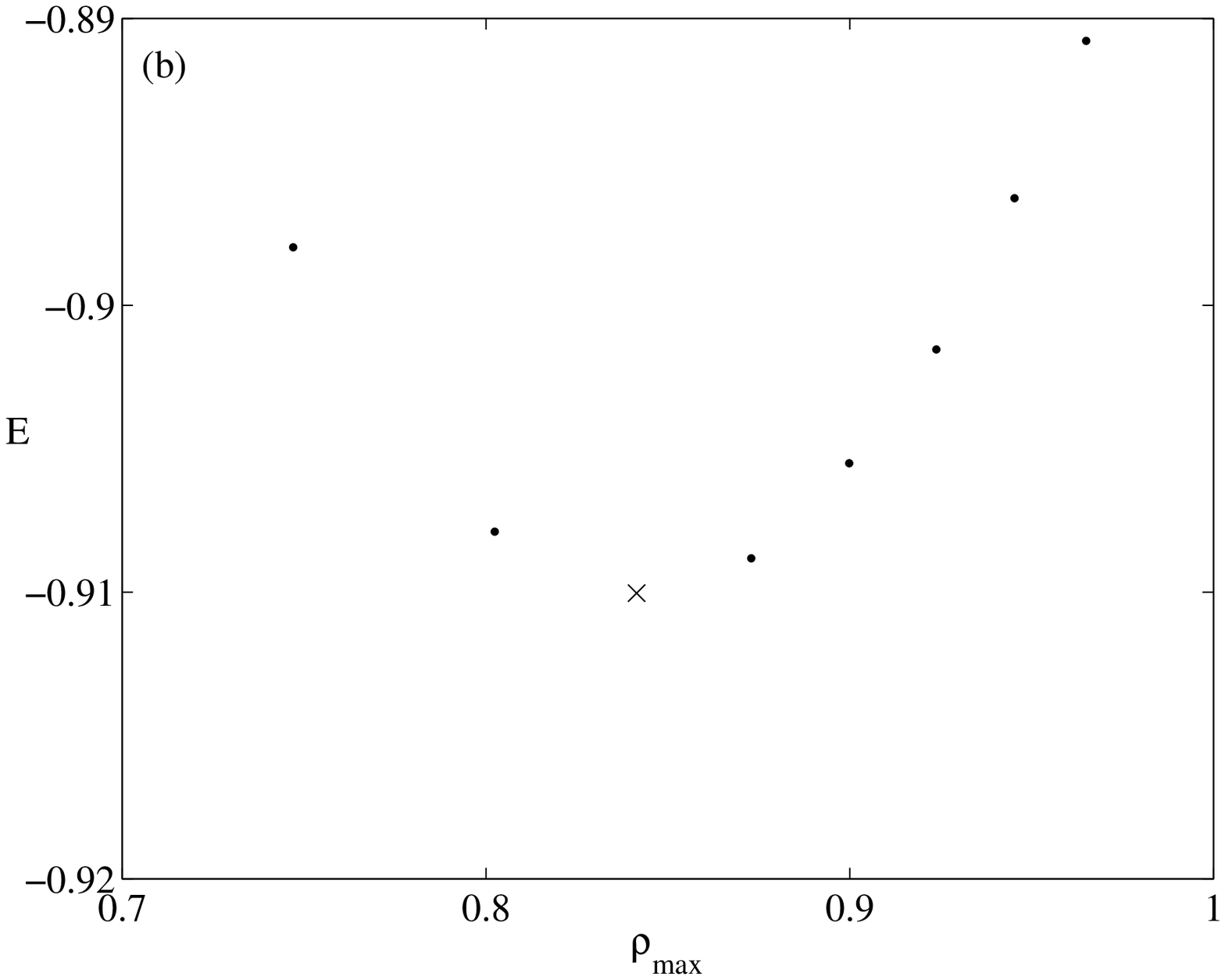}}}
\caption{(a) Members of the continuous family of compactly supported steady state solutions to (\ref{eq:ge}) having total population size $M=2.51$. Solutions are obtained from numerical integration of the phase-plane problem (\ref{eq:pp}). The thick broken line corresponds to the profile having the minimum energy E in (\ref{eq:energy}), which is the one observed in simulations of (\ref{eq:ge}). (b) Energies of the solutions in (a), parameterized by maximum density $\rho_{max}$. The minimum is indicated by an ``x''. For this example, $r=1$ in (\ref{eq:ge}).}
\label{fig:energy}
\end{figure}

As discussed in Section \ref{sec:pp}, for fixed population size $M$ there is a continuous family of clump solutions. However, as stated previously, and as demonstrated by the simulations of Section \ref{sec:coarsen}, one member of this family is preferred by the dynamics of (\ref{eq:ge}), and it is precisely the one which minimizes the energy (\ref{eq:energy}). Figure \ref{fig:energy} verifies this selection mechanism. Figure \ref{fig:energy}a shows some members of the continuous family, calculated from (\ref{eq:pp}) for $M=2.51$ with $r=1$ in (\ref{eq:ge}). The minimum-energy solution appears as a thick broken line. Figure \ref{fig:energy}b shows the energy $E$ of the solutions, parameterized by their maximum density $\rho_{max}$. To verify the energy argument, we evolve a random initial condition with $M=2.51$ according to (\ref{eq:ge}) using an implicit, central in space $(1,2)$-accurate finite difference method. When plotted on Figure \ref{fig:energy}a, the steady state achieved in the simulation is indistinguishable from the energy-minimizing solution in \ref{fig:energy}a calculated from (\ref{eq:pp}).

\begin{figure}
\centerline{\resizebox{\textwidth}{!}{\includegraphics{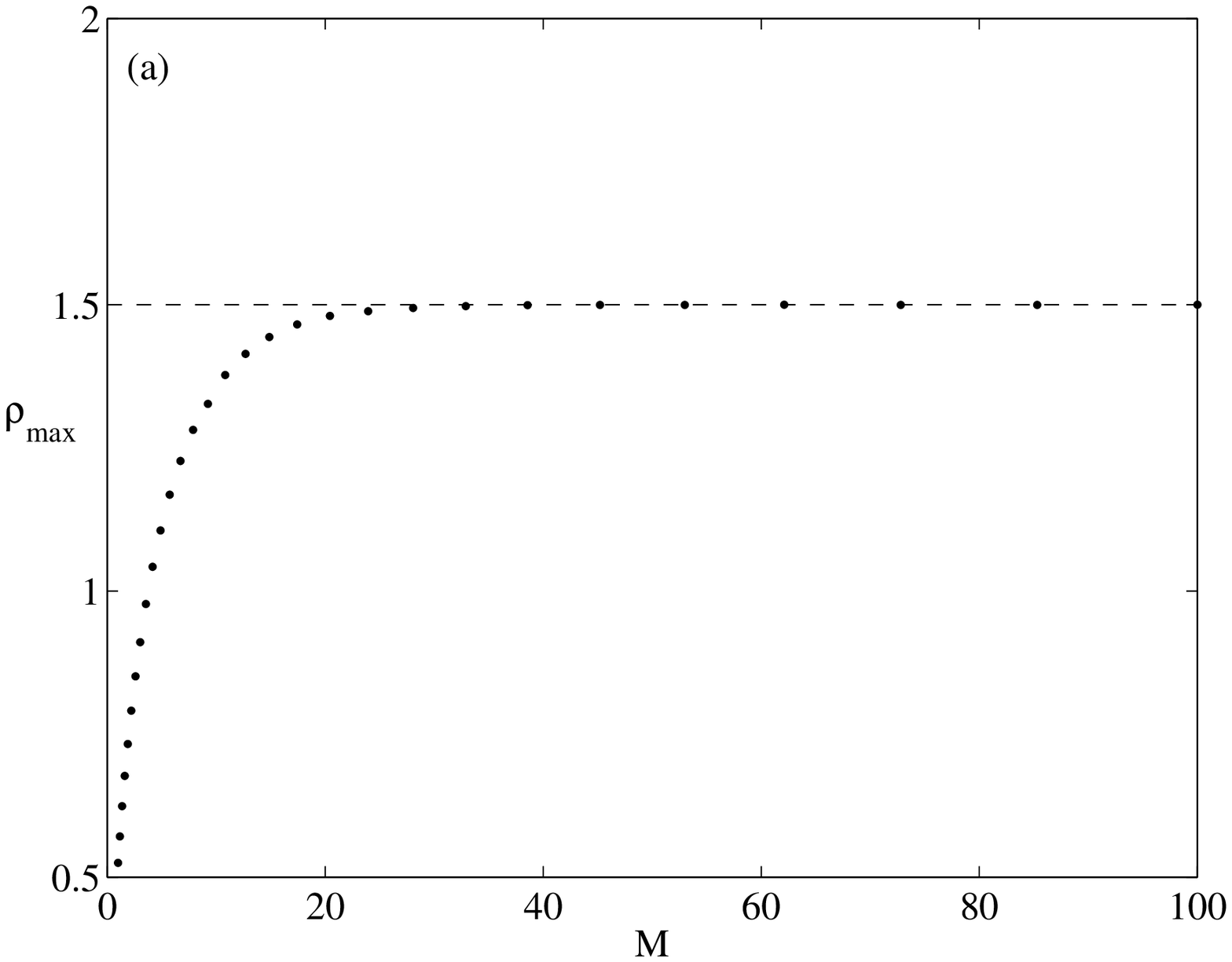} \hspace{0.1in} \includegraphics{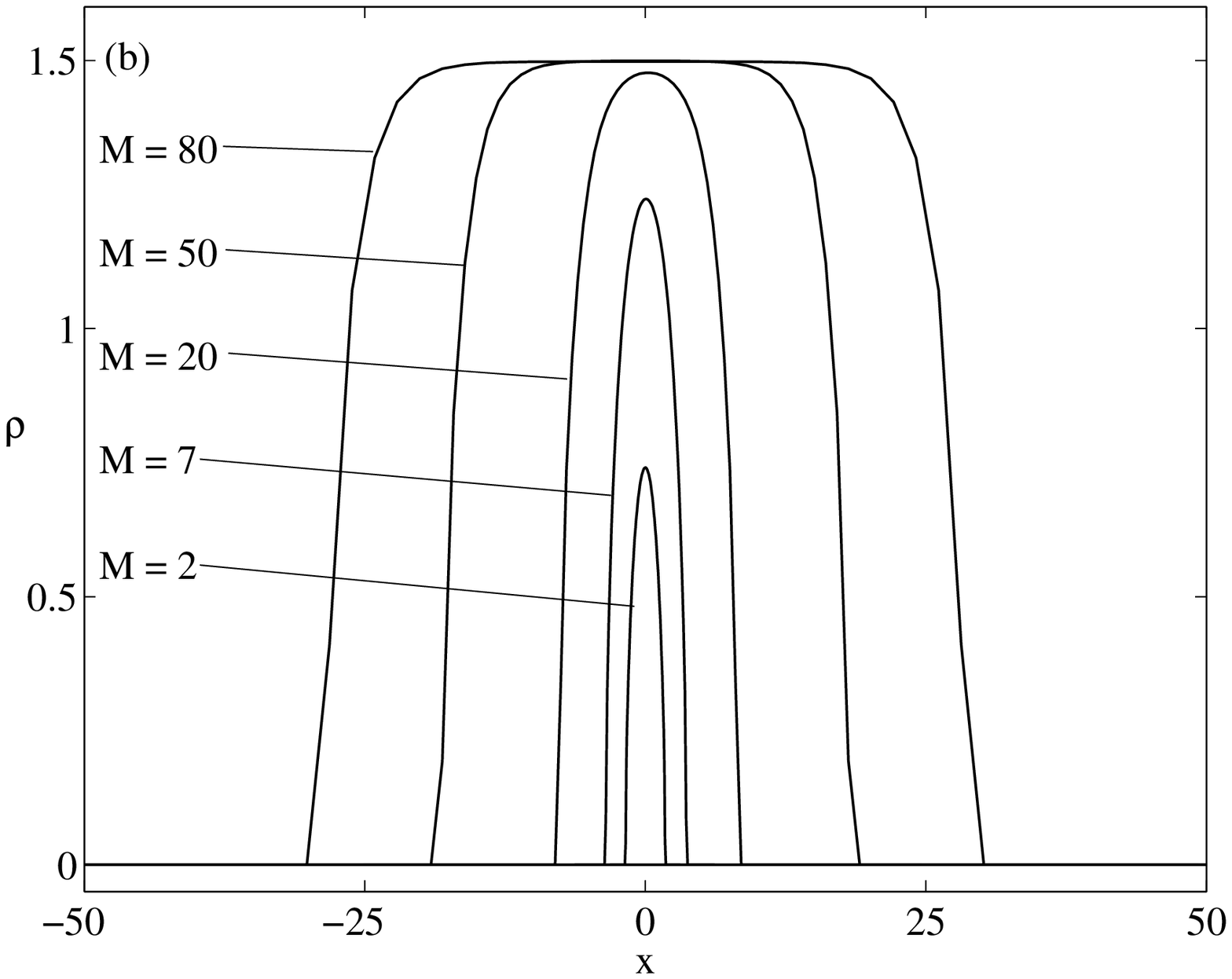}}}
\caption{Peak density $\rho_{max}$ versus population size $M$ of clump solutions computed on periodic domains of very large length $L$ so as to approximate solutions on $\mathbb{R}$ (see text for details). In the large $M$ limit, density profiles have $\rho_{max}=1.5$ as predicted by the energy (\ref{eq:energy}). (b) Density profiles for various values of $M$. Only the center of the domain, which contains the entire support of the clumps, is shown.}
\label{fig:clumps}
\end{figure}

Though thus far we have considered small periodic domains, it is perhaps more biologically relevant to consider the steady states of (\ref{eq:ge}) for the free space problem, \emph{i.e.} with $D=\mathbb{R}$. We approximate these solutions by taking an extremely large domain length $L$. We calculate the clump solutions for different population sizes $M$ with $L$ taken to be large enough for each $M$ so that the previously discussed finite-domain effect is negligible (a rule of thumb is that for a fixed $L$, we calculate only those clumps with $M \leq 0.5 M^c$). In Figure \ref{fig:clumps}a, we plot $\rho_{max}$ as a function of $M$. Selected density profiles are shown in Figure \ref{fig:clumps}b. For small values of $M$, \emph{e.g.} $M=2$ and $M=7$, the clumped solutions have small $\rho_{max}$ and a narrow, rounded shape. For larger population sizes, \emph{e.g.} $M=20$, $\rho_{max}$ is around $1.5$, and the shape of the profile is still quite rounded. For much larger population sizes, \emph{e.g.} $M=50$ and $M=80$, the clumps approach a rectangular shape, \emph{i.e.} they have a nearly constant internal density, deviating only very close to the edges of the group which go steeply to $\rho=0$. Interestingly, as $M$ is made larger and larger, $\rho_{max}$ stays at $1.5$ (see Figure \ref{fig:clumps}a) and the rectangular profile simply becomes wider.

The asymptote at $\rho_{max}=1.5$ in Figure \ref{fig:clumps}a may be understood from scaling and energy arguments. If we assume that the steady solution has a characteristic density $\rho_*$ and a characteristic length scale $L_*$, then the left side of the steady state equation (\ref{eq:ss}) scales as $\rho_*/L_*$, the first term on the right hand side scales as $\rho_*^2/L_*$, and the remaining term scales as $\rho_*^2/L_*^3$. In the limit of ``long'' solutions, the balance is between the left hand side and the first term on the right hand side, with the remaining term being a much smaller $\order(1/L^2)$ correction. Then we have $\rho_*/L_* \sim \rho_*^2/L_*$, and so $\rho_*$ is an $\order(1)$ constant. If we assume the population profile approaches a rectangular shape, we may use the energy (\ref{eq:energy}) to determine the height  $\rho_*$ of the rectangle. For rectangular solutions, (\ref{eq:energy}) yields
\begin{equation}
E = \int_{D} \frac{r}{3}\rho_*^3 - \rho_* K * \rho_*\ dx.
\end{equation}
$K$ has characteristic length scale $1$ (see Section \ref{sec:model}), and so for sufficiently large $L_*$, $K$ resembles a $\delta$-function and convolution with $K$ acts as the identity. Then we have
\begin{subequations}
\begin{eqnarray}
E & = & \int_{D} \frac{r}{3}\rho_*^3 - \rho_* ^2\ dx \\
& = &  L_*\left(\frac{r}{3}\rho_*^3 - \rho_* ^2\right) \\
& = & \frac{M}{\rho_*}\left(\frac{r}{3}\rho_*^3 - \rho_* ^2\right) \\
& = & M \left(\frac{r}{3}\rho_*^2 - \rho_*\right) \label{eq:selection}
\end{eqnarray}
\end{subequations}
where we recall that for a rectangular solution, $M=\rho_* L_*$. As previously discussed, for fixed $M$, the dynamically preferred steady state is the one which minimizes $E$. Minimizing (\ref{eq:selection}) with respect to $\rho_*$, we find $\rho_* = 3/(2r)$, in agreement with our numerical observations for $r=1$. From the inverse dependence of $\rho_*$ on $r$, we have the general result that the constant density of large rectangular groups is directly proportional to the ratio of characteristic attractive to repulsive velocities. Note that all of the above arguments are independent of the particular choice of interaction kernel $K$ after rescaling.

\section{Simulations in two spatial dimensions}
\label{sec:2d}

The energy argument given above is, in fact, independent of the number of dimensions, and so the preferred density for large clumps is always expected to be $\rho_* = 3/(2r)$. This is demonstrated by numerical simulations. We conduct our simulations on a periodic box. We use a hybrid numerical method with adaptive time stepping. The attraction term is treated explicitly and pseudospectrally. The dispersal term is treated implicitly, with operator splitting. For each time iteration, dispersal in the x and y directions are implemented successively (the order is alternated). The derivatives are approximated using central finite differences, and the resulting nonlinear problem is solved with Newton iteration.

\begin{figure}
\centerline{\resizebox{\textwidth}{!}{
\includegraphics{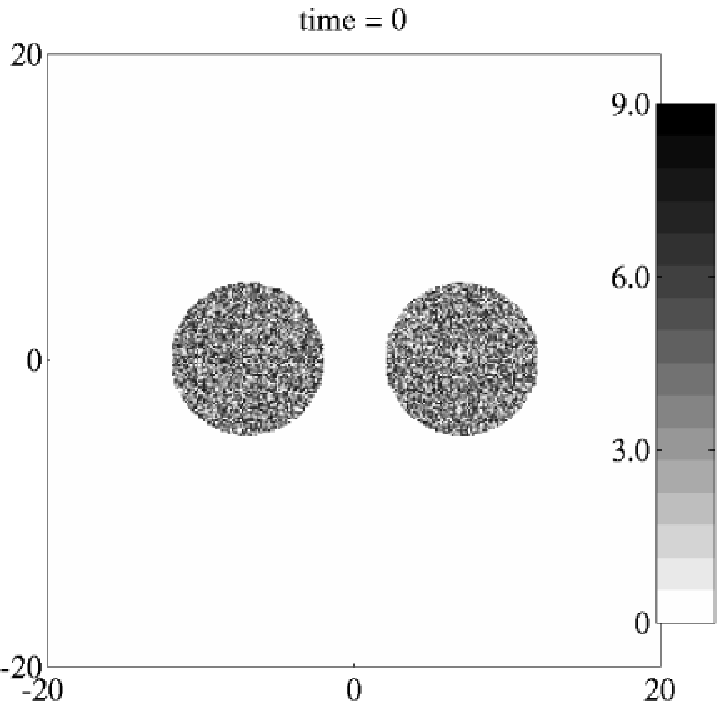} \hfill
\includegraphics{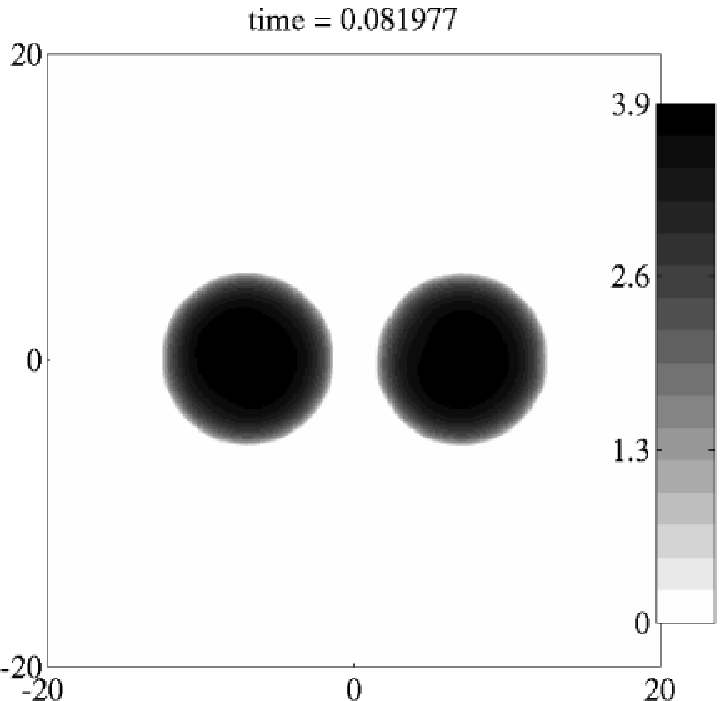}}}
\vspace{0.15in}
\centerline{\resizebox{\textwidth}{!}{
\includegraphics{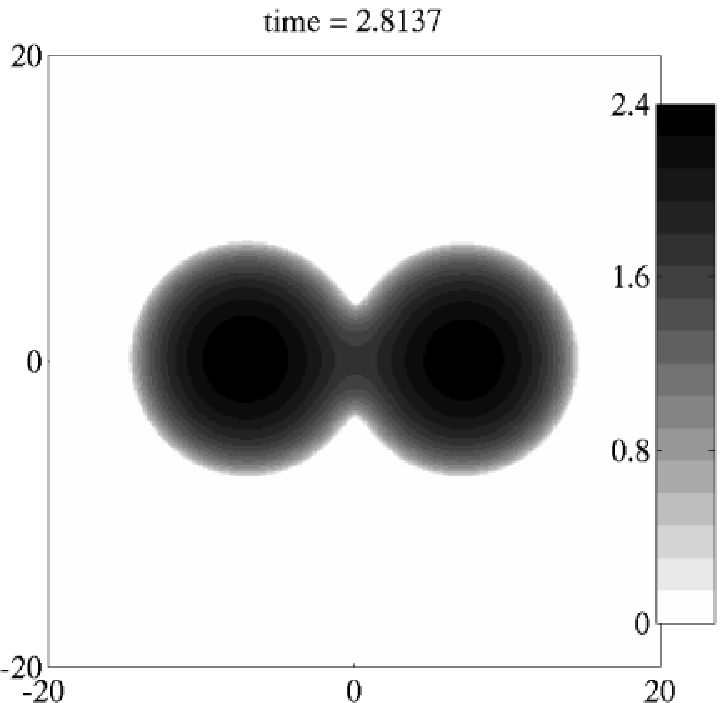} \hfill
\includegraphics{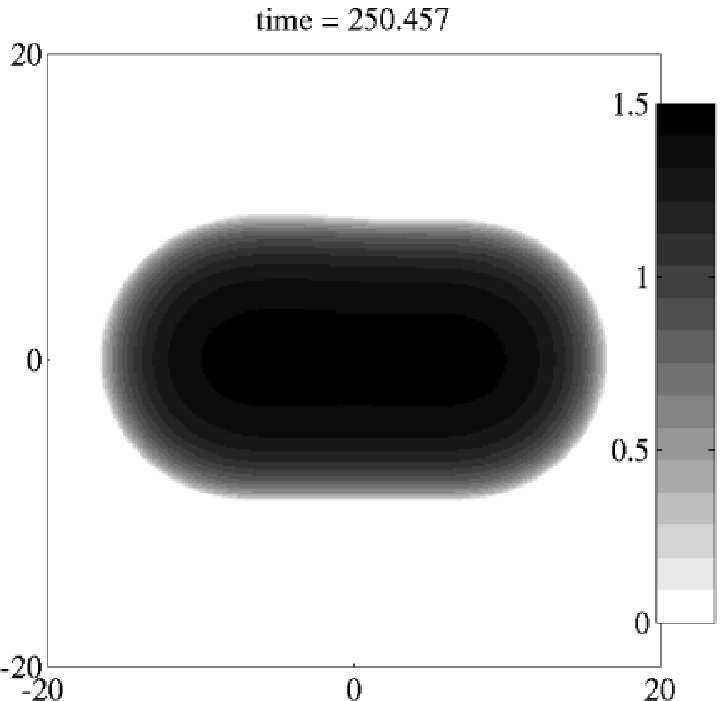}}}
\caption{(a-d) Population density profiles under the dynamics of (\ref{eq:ge}). The initial condition consists of two disjoint discs, with the population randomly distributed in each. The population in each disc rapidly smooths out, and the two discs merge over a slower time scale, and form a group with a nearly constant internal population density. The simulations are conducted with population size $M=600$ and characteristic repulsive to attractive speed ratio $r=1$ on a box of size $L=40$ with $N=300$ gridpoints on each axis.}
\label{fig:2d}
\end{figure}

\begin{figure}
\centerline{\resizebox{\textwidth}{!}{
\includegraphics{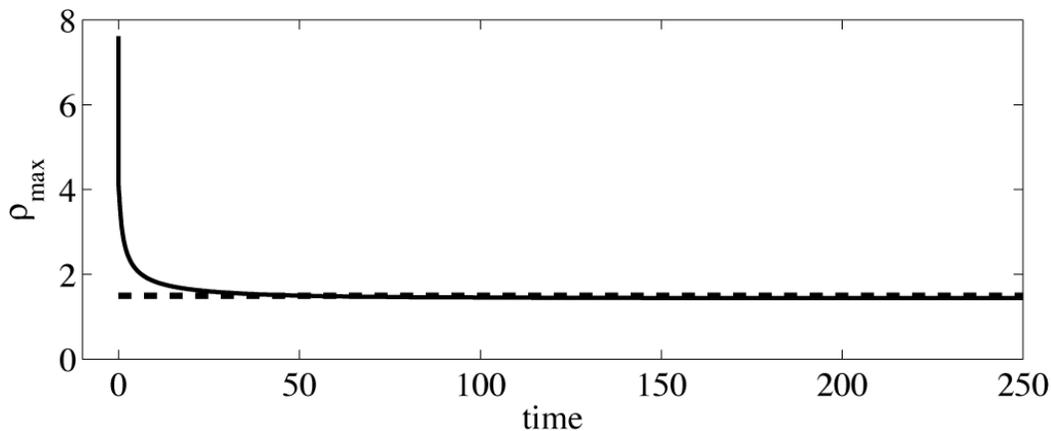}}}
\caption{Peak density $\rho_{max}$ as a function of time for the simulation shown in Figure~\ref{fig:2d}. The peak density approaches $1.5$, as predicted by energy arguments.}
\label{fig:2drhomax}
\end{figure}

Figure \ref{fig:2d} shows a numerical simulation of (\ref{eq:ge}) in a two dimensional box with each side $L=40$. The population size is $M=600$ and the speed ratio $r=1$. We take $N=300$ points along each axis. The initial condition (Figure \ref{fig:2d}a) consists of two disjoint discs, with a randomly distributed population in each one. The density profile within each disk rapidly smooths out (Figure \ref{fig:2d}b). Over a slower time scale, the two discs merge (Figure \ref{fig:2d}c). The resulting population clump retains a thin interfacial region outside of which $\rho=0$. The interface evolves on a very slow time scale, and the internal population density becomes nearly constant (Figure \ref{fig:2d}d). Because the evolution of the group boundary is very slow, we terminate the simulation before steady state is reached. Nonetheless, Figure \ref{fig:2drhomax} shows the peak density $\rho_{max}$ as a function of time, and confirms that it approaches $1.5$, as predicted by the energy arguments.

\section{Conclusion}
\label{sec:conclusion}

The purpose of this paper is to demonstrate how population clumping can occur in a realistic, minimal model for the movement of organisms. We have focused on a kinematic model, \emph{i.e.} one in which the velocity is a functional of the density (as opposed to a dynamic grouping model, which would include inertial forces). We have shown that the balance between nonlinear diffusion and nonlocal social attraction (which decays with distance) is sufficient for the formation of localized groups from initially disperse population profiles. Our focus has been on stationary, rather than traveling, solutions. Following the ideas in \citet{tb2004}, group motion may be incorporated into the model by incorporating additional social interaction terms of incompressible form, which have been shown to provide cohesive group motion while preserving the properties of constant density and compact support. Alternatively, group motion may be provided by considering external factors such as movement up food gradients. Of course, our model can also be interpreted simply as one for the formation of stationary clumps of organisms. 

Our case study in one spatial dimension uncovered three key properties which agree with biological observation:
\begin{enumerate}
\item{The groups are \emph{truly} localized, \emph{i.e.} they have well defined boundaries outside of which the population density is $0$.}
\item{The density drops steeply to $0$ at the edge, \emph{i.e.} the density profile has infinite slope at the boundary.}
\item{For sufficiently large populations, there is a preferred density (or in the parlance of discrete models, a preferred inter-organism spacing) independent of the population size itself, \emph{i.e.} adding more members to the group does not change the ``packing'' of organisms, but simply increases the spatial extent of the group.}
\end{enumerate}

Furthermore, for the case of two spatial dimensions, simulations revealed solutions which also have compact support, steep edges, and a constant internal population density. Finally, we have connected a macroscopic property of the population clumps, namely the preferred internal population density, to a property of the movement rules, namely the ratio of typical dispersive to attractive speeds, by analyzing an energy associated with the aggregation.

\ack

The research of CMT is supported by NSF VIGRE grant DMS-9983726. ALB is supported by ARO grant DAAD-19-02-1-0055, ONR grants N000140310073 and N000140410054, and NSF grant DMS-0244498. MAL is supported by grants from NSERC and the Canada Research Chairs program. We are grateful to Andrew Bernoff, Dejan Slepcev and Tom Witelski for helpful discussions.

\bibliographystyle{elsart-harv}
\bibliography{bibliography}

\end{document}